\documentclass[superscriptaddress, dvipsnames, amsmath, amssymb, aps, prx, twocolumn, longbibliography]{revtex4-1}
\usepackage{graphicx}
\usepackage{dcolumn}
\usepackage{bm}
\usepackage{footnote}
\usepackage{mathtools}
\usepackage{psfrag}
\usepackage{xcolor}
\usepackage{amsthm}

\usepackage[colorlinks = true, allcolors = blue]{hyperref}
\usepackage{dsfont}
\usepackage{tabularx}
\usepackage[capitalize]{cleveref}
\usepackage{color}
\usepackage{booktabs}

\begin{document}

\title{Work extraction with feedback control using limited resources}

\author{Harrison Hartle}
\email{hthartle1@gmail.com}
\affiliation{Santa Fe Institute, Santa Fe, New Mexico 87501, United States} 
\author{David Wolpert}
\affiliation{Santa Fe Institute, Santa Fe, New Mexico 87501, United States} 
\author{Andrew J. Stier}
 \affiliation{Santa Fe Institute, Santa Fe, New Mexico 87501, United States} 
\author{Christopher P. Kempes}
\affiliation{Santa Fe Institute, Santa Fe, New Mexico 87501, United States} 
\author{Gonzalo Manzano}
\email{gonzalo.manzano@ifisc.uib-csic.es}
\affiliation{Institute for Cross-Disciplinary Physics and Complex Systems (IFISC) UIB-CSIC, Mallorca, Spain}

\date{\today}
            
\begin{abstract} 
Many physical, biological, and even social systems  are faced with the problem of how to efficiently harvest free energy from an environment that can have many possible states, yet only have a limited number of harvesting protocols to choose among. We investigate this scenario by extending earlier work on using feedback control to extract work from nonequilibirum systems. Specifically, in contrast to that previous work on the thermodynamics of feedback control, we analyze the combined and separate effects of noisy measurements, memory limitations, and limitations on the number of possible work extraction protocols. Our analysis provides a general recipe to construct repertoires of allowed harvesting protocols that minimize the expected thermodynamic losses during free energy harvesting, i.e., that minimize expected entropy production. In particular, our results highlight that the benefits of feedback control over uninformed (random) actions extend beyond just the associated information gain, often by many orders of magnitude. Our results also uncover the effects of limitations on the number of possible harvesting protocols when there is uncertainty about the distribution over states of the environment. 
\end{abstract}

\maketitle

\section{Introduction}

A central concern of essentially all living organisms is that they must interact intelligently with their nonequilibrium environment in order to extract enough free energy from it to survive. This in turn requires them to acquire information concerning the state of their environment via measurements, to guide their behavior in those interactions~\cite{Bialek2016}. Relatedly, a long line of scientists have been fascinated by the interplay between the probabilistic nature of the second law and the possibility to use information to improve thermodynamic operation~\cite{leff2002maxwell} since Maxwell's demon thought experiment in the 19th century and the first information engine using feedback control operations proposed by Leo Szilard~\cite{Szilard1929}.

The development of stochastic thermodynamics in the last decades has provided the possibility of a precise formulation and analysis of feedback control scenarios~\cite{parrondo2015thermodynamics} where the costs and benefits of using information can be assessed in a variety of situations, ranging from discrete measurements~\cite{Touchette2000,Sagawa2009,Sagawa2010,Horowitz2010,Abreu2012}, to continuous monitoring~\cite{Sagawa2012,Deffner2013,Horowitz2013,Potts2018,Ribezzi2019,Debiossac2020}, and from classical to quantum systems~\cite{Sagawa2008,Funo2013,Francica2017,Manzano18L,Annby2022}. Experiments in a broad range of platforms from colloidal particles~\cite{Toyabe2010} and nanoelectronic devices~\cite{Koski2014P,Koski2014} to DNA pulling experiments~\cite{Rico-Pasto2021} have tested and verified the role of information in work extraction. 

In this paper we focus on using feedback protocols to guide the extraction of work (i.e., the harvesting of free energy) from a nonequilibrium system of interest (e.g. the environment of an organism). We consider specifically the effects arising from noisy observation and from a limited number of possible actions on the nonequilibrium system, as well as limitations on the size of the memory used in the feedback control operation. Note in particular that limitations on the number of possible actions by the agent extracting work apply in many real-world scenarios. For example, they arise in switching on and off predefined potentials~\cite{Toyabe2010}, suddenly raising the energy of a discrete level~\cite{Koski2014P} or swapping the polarity of photons~\cite{Vidrighin2016}. More generally, such limitations apply to all biological organisms that are trying to harvest free energy from their environment.

Here we show that such limitations can drastically reduce the efficiency of free energy extraction performed via feedback control. As a result, determining the optimal finite set of actions the controller can choose among, given such memory limitations and noise in measurements is crucial in several applications. These include cases where the initial distribution of the system is not known and might be guessed from noisy observation.
To analyze these issues, here we build on previous work demonstrating that optimal feedback control which maximizes work extraction can be achieved with reversible feedback protocols~\cite{Horowitz2011,Horowitz2011b}. Such optimal
control begins with a quench of the original Hamiltonian of the nonequilibrium system.
The precise quench depends on the original distribution of system states and any (noise-free) measurement concerning that state. After this quench a quasi-static isothermal process is applied to the system, changing the Hamiltonian back to the original Hamiltonian, and changing the distribution to an equilibrium distribution with respect to that original Hamiltonian~\cite{Hasegawa2010,Takara2010,Esposito2011,Anders2013,parrondo2015thermodynamics}.

However, as mentioned, in real-world scenarios limitations often arise due to both imperfections in the measurements and because of severe constraints in the control operations that prevent the implementation of optimal protocols~\cite{Bechoefer2024}. For example, if the relaxation timescales of the system are not fast enough with respect to the driving velocity, the quasi-static regime is lost. In that case, optimal transport techniques are required to maximize the extractable work that usually involve numerical techniques~\cite{Aurell2011}. In some cases, however, the problem can be directly solved in a linear response regime if the dynamics is sufficiently slow (i.e. close to but out of the quasi-static regime)~\cite{Sivak2012,Zulkowski2012,Proesmans2020,Scandi2022}. Moreover, in many situations, the dynamical evolution of a system may show special symmetry or modularity properties that can affect its thermodynamics. In such cases, extra irreversibility may arise that can limit work extraction even when allowing generic thermodynamic processes under local detailed balance~\cite{Kolchinsky2021,Kolchinsky2021E}. 

Here we will assume for simplicity no limitations associated to the relaxation time-scales, so that quasi-static processes are feasible, and no extra symmetries. We moreover focus on feedback control processes mimicking the optimal ones, i.e., consisting in the implementation of a sudden quench on the system Hamiltonian that depends on the measurement result, followed by quasi-static isothermal evolution that takes the Hamiltonian back to its original form. All the limitations considered here hence regard the set of available quenches that can be performed to extract work from the initial (nonequilibrium) state of the system by transforming it into an equilibrium state (with respect to the quenched Hamiltonian). Also, in many real-world scenarios, the experimentalist has little (if any) direct control over rate matrices, beyond that arising via local detailed balance and their ability to change a system's energy spectrum. Accordingly (and in contrast to~\cite{Kolchinsky2021,Kolchinsky2021E}), we assume no control of rate matrices whatsoever beyond that which arises automatically due to local detailed balance.

Our main results involve the derivation of universal expressions for the entropy production arising due to the control limitations, which limit the amount of extractable work (free energy) from generic nonequilibrium states. The general framework is outlined in Sec.~\ref{sec:framework} starting from the ideal case and then introducing control limitations. That framework allows us to obtain in Sec.~\ref{sec:information} precise conditions for the design of successful feedback control quenches that balance the eventual work costs with a net free energy extraction. There we also determine the quantitative benefits of employing information with respect to scenarios where the quenching Hamiltonian is fixed or taken at random from a predefined set. In Sec.~\ref{sec:optimal_extraction} we derive explicit expressions for the optimal form of the quenches that minimize entropy production and present our results for the construction of an optimal action repertoire. We discuss its implications in terms of the acquired information that is useful for feedback control. Our general results are particularized in various situations of interest in stochastic thermodynamics in Sec.~\ref{sec:applications}. Noise-free observation and observation with small errors are considered, as well as the case of a random initial distribution, all of which illustrated using worked examples. Finally, in Sec.~\ref{sec:conclusions} we provide a concluding summary and discussion. Details about the derivations of the main results are given in Appendices \ref{app:optimal_quench_given_partition}, \ref{app:perfect_observation}, \ref{app:1D}, and \ref{app:random_initial}, and Table~\ref{TABLE} is a guide to the notation used in the manuscript.

\section{Framework}
\label{sec:framework}

Consider a non-equilibrium system of interest from which work is to be extracted. It is described by a set of $d_X$ variables $\mathbf{x}:= \{x_i \}_{i=1}^{d_X}$ (we use bold letters to indicate vectors), with $x_i \in \mathbb{X}$, where the set $\mathbb{X}$ can be either discrete or continuous. We introduce an associated Hamiltonian function describing the energy of the system over this set of variables $H(\mathbf{x})$ which can be externally controlled by an external agent subjected to computational limitations that will be specified in detail later. The system is assumed to be initially in a non-equilibrium state described by distribution  $\rho_0(\mathbf{x})$. 

A feedback controller can extract work from the nonequilibrium state of the system by first measuring (all or some of) the system variables, and then performing a control operation over the system Hamiltonian conditioned on the obtained measurement result. 
The measurement provides the controller with a vector of measurement results $\mathbf{m}:= \{ m_i\}_{i=1}^{d_M}$ with $m_i \in \mathbb{M}$ (which can also be either discrete or continuous). The measurement results are affected by external noise as described by the (joint) conditional probability $p(\mathbf{m} | \mathbf{x})$ for obtaining outcome $\mathbf{m}$ given that the state of the system was $\mathbf{x}$. The (marginal) probability of obtaining the results $\mathbf{m}$ is therefore a function of the initial state $\rho_0(\mathbf{x})$ and the above conditional probabilities as given by 
\begin{equation}
 p(\mathbf{m}) = \sum_{ \mathbf{x}} p(\mathbf{m} |\mathbf{x}) \rho_0(\mathbf{x}). 
\end{equation}
After measurement, the best guess for the system distribution hence reads:
\begin{align} \label{eq:Rhom}
  \rho_{X|\mathbf{m}}(\mathbf{x}) := \rho(\mathbf{x} | \mathbf{m}) = \frac{p(\mathbf{m}| \mathbf{x}) \rho_0(\mathbf{x})}{p(\mathbf{m})} 
\end{align}
as corresponds to Bayes' rule for updating probabilities when using $\rho_0(\mathbf{x})$ as the prior distribution. Note that we use $\rho_{X|\mathbf{m}}(\mathbf{x})$ as a short-hand notation to denote the conditional distribution $\rho(\mathbf{x} | \mathbf{m})$ for a given measurement result $\bold{m}$. Analogously we will use in the following $p_{M|\bold{x}}(\mathbf{m})$ to denote the conditional probability $p(\mathbf{m} |\mathbf{x})$.
In the case of a perfect measurement of the system variables (i.e., the measurement result $m_i$ corresponds exactly to the system state variable $x_i$ for all $i$) the above expressions reduce to $p_{M|\bold{x}}(\mathbf{m}) = \prod_{i=1}^d \delta_{m_i, x_i}$, where $\delta_{m,x}$ denotes the Kronecker delta, and hence $p(\mathbf{m}) = \rho_0(\mathbf{m})$, implying that the state of the system becomes perfectly known, i.e., $\rho_{X|\mathbf{m}} (\mathbf{x})=\prod_{i=1}^d \delta_{m_i, x_i}$. See Sec.~\ref{ssec:perfect_observation} for how our main results apply in this setting.

\subsection{Ideal work extraction}
The maximum amount of work that can be extracted from the system, assuming knowledge of $\rho_{X|\mathbf{m}} (\mathbf{x})$, and access to an equilibrium reservoir at temperature $T$, is the difference in nonequilibrium free energies between such a state and the corresponding equilibrium state $\rho_H^{\rm eq}(\mathbf{x})= e^{- \beta H(\mathbf{x})}/Z$ with $Z = \sum_{\mathbf{x}} e^{- \beta H(\mathbf{x})}$ being the partition function, and $\beta = 1/(k_B T)$, where $T$ is the temperature of the thermal reservoir connected to the system (or bath) ~\cite{parrondo2015thermodynamics}. That is:
\begin{equation} \label{eq:Wext}
 W(\mathbf{m}) \leq \mathcal{F}_H(\rho_{X|\mathbf{m}}) - F_H^{\rm eq},
\end{equation}
where the nonequilibrium free energy is defined as the function:
\begin{equation}
 \mathcal{F}_H(\rho):= \langle H \rangle_\rho - k_B T S(\rho),
\end{equation}
with $\langle H \rangle_\rho = \sum_{\mathbf{x}} H(\mathbf{x}) \rho(\mathbf{x})$ the average energy of the system associated to distribution $\rho$ and $S(\rho) = -\sum_{\mathbf{x}} \rho(\mathbf{x}) \ln \rho(\mathbf{x})$ its Shannon entropy. Notice that for equilibrium states $\mathcal{F}_H(\rho_H^{\rm eq})$, the nonequilibrium free energy reduces to the traditional equilibrium (Helmholtz) free energy for Hamiltonian $H$, which we denote $F_H^{\rm eq}$.

The feedback control protocol achieving optimal work extraction [equality in Eq.~\eqref{eq:Wext}] can be constructed in a generic way by first instantaneously quenching the Hamiltonian to
\begin{equation} \label{eq:ideal-quench}
H(\mathbf{x}) \rightarrow H_\mathbf{m}(\mathbf{x}) := -k_B T \ln \rho_{X|\mathbf{m}}(\mathbf{x}), 
\end{equation}
and then quasi-statically modifying from $H_\mathbf{m}(\mathbf{x})$ back to the original Hamiltonian $H(\mathbf{x})$, while in contact with the external bath~\cite{parrondo2015thermodynamics}. Recall that the rationale behind this procedure follows from the fact that the initial Hamiltonian quench instantaneously transforms the state $\rho_{X|\mathbf{m}}(\mathbf{x})$ into being an equilibrium state with respect to the quenched Hamiltonian $H_\mathbf{m}(\mathbf{x})$, after which the quasi-static relaxation step just becomes a thermodynamically reversible transformation where the state of the system is, at every time, in equilibrium with the bath. By following this procedure, all of the available free energy is extracted.  We write the work extracted using this (ideal) feedback control protocol as:
\begin{align} \label{eq:Ext}
 & W_{\rm ideal}(\mathbf{m}) = -\langle H_\mathbf{m} \rangle_{\rho_{X|\mathbf{m}}} + \langle H \rangle_{\rho_{X|\mathbf{m}}} \nonumber \\ &  + \mathcal{F}_{H_\mathbf{m}}(\rho_{X|\mathbf{m}}) - F_H^{\rm eq} = \mathcal{F}_H(\rho_{X|\mathbf{m}}) - F_H^{\rm eq}.
\end{align}
This formula follows from the fact that the free energy of the quenched state is $\mathcal{F}_{H_\mathbf{m}}(\rho_\mathbf{m}) = \langle H_{\mathbf{m}} \rangle_{\rho_{X|\mathbf{m}}} - k_B T S(\rho_{X|\mathbf{m}})$ and that during the reversible transformation the work extracted equals the (equilibrium) free energy change.

Notice that $\mathcal{F}_H(\rho_{X|\mathbf{m}})$ is a stochastic quantity depending on the specific outcomes obtained in the measurement, and hence so is $W_{\rm ideal}(\mathbf{m})$. The associated average over measurement results is:
\begin{align} \label{eq:Wextav}
 \langle W_{\rm ideal} \rangle &= \sum_{\mathbf{m}} p(\mathbf{m}) \mathcal{F}_H(\rho_{X|\mathbf{m}}) - F_H^{\rm eq} \nonumber \\ &= \langle H \rangle_{\rho_0} - k_BT \sum_{\mathbf{m}} p(\mathbf{m})S(\rho_{X|\mathbf{m}}) -  F_H^{\rm eq}  \nonumber \\
 &= k_BT I_{X; M} + \mathcal{F}_H(\rho_{0}) - F_H^{\rm eq} ,
\end{align}
where we identified $I_{X;M} := S(\rho_0) -  \sum_{\mathbf{m}} p(\mathbf{m}) S(\rho_{X|\mathbf{m}})$ as the mutual information between the system variables and measurement outcomes. In the equation above we used that the unconditional state of the system after measurement is just the initial state, $\sum_{\mathbf{m}} p(\mathbf{m}) \rho_{X|\mathbf{m}}(\bold{x}) = \rho_0(\bold{x})$, which can be directly verified from the expression of $\rho_{X|\mathbf{m}}$ in Eq.~\eqref{eq:Rhom} (a property of non-invasive measurements). Equation~\eqref{eq:Wextav} tells us that the amount of work extractable from the environment is limited by two factors (i) the amount of information that can be gathered from measuring the system of interest as given by the mutual information $I_{X;M}$ and (ii) the amount of free energy contained in the initial non-equilibrium state of the environment $\rho_0(\mathbf{x})$ as compared to its equilibrium free energy, which we denote in the following $\Delta \mathcal{F}_H := \mathcal{F}_H(\rho_0) - F_H^{\rm eq}$.

It is worth mentioning at this point that closing the cycle after the work extraction procedure requires a final step consisting in the erasure of the information gathered during the measurement of the system and employed for the implementation of the feedback control protocol~\cite{parrondo2015thermodynamics}. Following generalized Landauer's principle, this erasure process has an unavoidable work cost that amounts, at least, to the mutual information $k_B T I_{X;M}$ in the reversible limit~\cite{Sagawa2009}. That means that the net effect of the work extraction procedure, once erasure costs of the memory have been included, is to transform the free energy difference from the initial nonequilibrium state $\rho_0(\mathbf{x})$ to the equilibrium one $\rho_H^{\rm eq}(\mathbf{x})$ into useful work.

\subsection{Introducing control limitations}

In contrast to the ideal work extraction scenario described above, in many real-world scenarios the set of possible quenching Hamiltonians is finite and much smaller than the set of possible outcomes of the measurement. In addition, often in the real world that set is predefined and fixed ahead of time, and the only thing one can do is choose one of those fixed possibilities. We are interested in quantifying how much the ideal scenario presented above is affected when introducing such limitations affecting the control abilities of the agent performing work extraction. 

More specifically, we consider a broad range of real-world scenarios characterized by the impossibility of the controller to perform the ideal quench introduced in Eq.~\eqref{eq:ideal-quench}, that leads to a post-quench Hamiltonian $H_\mathbf{m}(\mathbf{x})= -k_B T \ln \rho_{X|\mathbf{m}}(\mathbf{x})$. Instead, we consider the situation in which the feedback controller action is limited to choose one out of some finite set of quench Hamiltonians. We write that set as $\mathcal{H} := \{H_k(\mathbf{x}) \}_{k=1}^N$ with $N <|\mathbb{M}^{d_M}|$ possible options. In this situation, after reading the measurement results $\mathbf{m}$, the feedback controller selects and applies one of the quenches in the repertoire $\mathcal{H}$ so as to maximize the amount of extractable work, according to the obtained measurement result. We denote the choice by the control protocol of
which Hamiltonian to quench to in response to a given measurement result $\mathbf{m}$ by $n(\mathbf{m})$, or $n$ for short. (We 
derive the optimal form of that function below.) Note that because
$\mathbf{m}$ is random, in general so will $n$ be. We will first assume the form of the elements $H_k(\mathbf{x})$ in $\mathcal{H}$ to be predefined and fixed, while in following sections we will also determine their optimal form.

In any case, as a consequence of the limitations in the available actions of the controller for the implementable (post-quench) Hamiltonians, the state of the system after the quench will no longer be in equilibrium with the post-quench Hamiltonian. In other words, $\rho_{X|\mathbf{m}}(\mathbf{x}) \neq -k_B T \ln H_{n(\bold{m})}(\mathbf{x})$. 

Hence an irreversible relaxation from the (initial) state $\rho_{X|\mathbf{m}}(\mathbf{x})$ to the actual equilibrium state after the quench $e^{-\beta H_n(\mathbf{x})}/Z_n$ (with $Z_n= \sum_x e^{-\beta H_n(\mathbf{x})}$) will unavoidably take place before the quasi-static process completing the work extraction protocol is completed. In this case the work extractable from the initial state using feedback control is reduced:
\begin{align} \label{eq:ImperfectW}
    W(\mathbf{m}) &= - \langle H_n \rangle_{\rho_{X|\mathbf{m}}} + \langle H \rangle_{\rho_{X|\mathbf{m}}} + F_{H_n}^{\rm eq}  - F_H^{\rm eq}  \\
    &= \mathcal{F}_H(\rho_{X|\mathbf{m}}) - F_H^{\rm eq}  - k_B T D\left({\rho_{X| \mathbf{m}}} \left\vert\left\vert \frac{e^{-\beta H_{n}}}{Z_{n}}\right.\right.\right). \nonumber
\end{align}
In the last equation, the first two terms account for the free energy extractable as work in the ideal case [c.f. Eq.~\eqref{eq:Ext}], where exact quenching to $H_{\mathbf{m}}(\mathbf{x}) = - k_B T \ln \rho_{X|\mathbf{m}}(\mathbf{x})$ is possible (and hence $Z_{\mathbf{m}} = 1$). The second term corresponds to the entropy production (or dissipation) generated in the irreversible post-quench relaxation $\rho_{X|\mathbf{m}}(\mathbf{x}) \rightarrow e^{-\beta {H}_n(\mathbf{x})}/Z_n$,
which we can write as
\begin{equation} \label{eq:Stot}
S_\mathrm{tot}(\mathbf{m}) =  D\left({\rho_{X|\mathbf{m}}} \left\vert\left\vert \frac{e^{-\beta H_{n}}}{Z_{n}}\right.\right.\right), 
\end{equation}
i.e., the Kullback-Leibler divergence between initial and final system densities~\cite{kawai2007dissipation}. (Recall that the Kullback-Leibler divergence $D(\rho || \sigma) \geq 0$ is well defined if $\rho$ has support on $\sigma$, it is non-negative $\forall \rho, \sigma$ and zero if and only if $\rho = \sigma$~\cite{CoverThomasBook}.)

We can hence rewrite Eq.~\eqref{eq:ImperfectW} as $W(\mathbf{m}) = W_\mathrm{ideal}(\mathbf{m}) - k_B T S_\mathrm{tot}(\mathbf{m})$. It is worth noticing that the first term, $W_\mathrm{ideal}(\mathbf{m})$, does not depend on the actual quenching Hamiltonian chosen in the imperfect scenario $H_n(\mathbf{x})$. It is only the second, entropy production term $S_\mathrm{tot}(\mathbf{m})$ that depends on that Hamiltonian. Note in particular that if the two distributions in that second term are very different (as quantified by their KL-divergence), entropy production may be large enough that the net work extracted in the process is negative. More explicitly, that will be the case whenever 
\begin{equation}
D \left({\rho_\mathbf{m}} \left\vert\left\vert \frac{e^{-\beta H_{n}}}{Z_{n}}\right.\right.\right) \geq \mathcal{F}_H(\rho_{X|\mathbf{m}}) - \mathcal{F}_H(\rho_H^{\rm eq})
\end{equation}
This shows that work might be \emph{lost} in the work extraction process, if the quenching Hamiltonian $H_n(\mathbf{x})$ is not adequately chosen.

Repeating this entire process many times and taking the average of the work extracted in  Eq.~\eqref{eq:ImperfectW} over the different measurement results, we obtain the average work extracted under control limitations:
\begin{align} \label{eq:avwork}
    \langle {W} \rangle =&~ k_B T I_{X;M} + \Delta \mathcal{F}_H 
    - k_B T  \langle S_\mathrm{tot} \rangle,
\end{align}
 where we used the short-hand notation $\Delta \mathcal{F}_H := \mathcal{F}_H(\rho_0) - F_H^{\rm eq}$. In the above equation the first two terms give the work extracted in the ideal case [c.f. Eq.~\eqref{eq:Wextav}] and the third one gives the average entropy production due to imperfect quenching, which reduces the amount of work extracted. That average entropy production over measurement results is given by a a related calculation:
\begin{equation} \label{eq:averageEP}
 \langle S_\mathrm{tot} \rangle  = \sum_\mathbf{m} p(\mathbf{m}) D\left(\rho_{X|\mathbf{m}} \left\vert\left\vert \frac{e^{- \beta H_{n(\mathbf{m})}}}{Z_{n(\mathbf{m})}}\right.\right.\right).
\end{equation}
These two calculations will be the basis of our main results. Again, erasure of the information employed for the feedback control in preparation for a next iteration of the entire process will have an unavoidable cost of $k_B T I_{X;M}$ imposed by the generalized Landauer's bound~\cite{parrondo2015thermodynamics}.

\section{The value of information under control limitations}
\label{sec:information}

Given the general scenario described above for quasi-static work extraction using a limited set of feedback control quenches, we will first compare the situation of a prefixed set of quenching Hamiltonians $\mathcal{H} = \{H_k(\mathbf{x}) \}_{k=1}^N$, where the controller is able to measure the system of interest and decide which action to take, to the case when no such informated choice can be made, and the controller may just apply either a deterministic Hamiltonian or perform a random choice in $\mathcal{H}$.

To make things more precise, we start by considering a partition of the entire space $\mathbb{M}^{d_M}$ of measurement results $\mathbf{m}$ into $N$ disjoint sets ${L} := \{l(n)\}_n$ with $n=1, ..., N$, such that $\cup_{n}\ell(n)=\mathbb{M}^{d_M}$ and $\ell(n)\cap\ell(n')=\varnothing$ for $n\ne n'$. In that way, every measurement result $\mathbf{m}$ can be unequivocally associated to a set $l(n)$ with label $n$. The probability that the measurement outcome $\mathbf{m}$ is inside the set $l(n)$ can 
then be written as:
\begin{equation}
    p(n):=\sum_{\mathbf{m} \in l(n)} p(\mathbf{m}).
\end{equation}
Notice that by this point we haven't impose any property on the size of the different sets $l(n)$.

We then ascribe one of the possible quenching Hamiltonians $H_n(\mathbf{x}) \in \mathcal{H}$ to each set $l(n) \in {L}$. Since we are considering a finite partition, the computation of $l(n)$ that associates the Hamiltonians in the set $\mathcal{H}$ with the labels $n$ (linked to the observations $\mathbf{m}$) is just a lookup table. This table returns the Hamiltonian $H_n(\mathbf{x})$ from the repertoire $\mathcal{H}$ verifying that $e^{-\beta H_n(\mathbf{x})}/Z_n$ is most similar to $\rho_{X|\mathbf{m}}(\mathbf{x})$ as quantified by Kullback-Leibler (KL) divergence. 
Formally, the optimal lookup table is the function
\begin{equation} \label{eq:argmin}
   n^*(\mathbf{m}) =\underset{k \in \{1,...,N\}}{\mathrm{argmin}}~ D\left({\rho_{X|\mathbf{m}}} \left\vert\left\vert \frac{e^{-\beta H_k}}{Z_k}\right.\right.\right). 
   \end{equation}
   In this way the labels $n$ associated to the Hamiltonians $H_n(\mathbf{x})$ in $\mathcal{H}$ are determined by the minimization of the entropy production for given outcome $\mathbf{m}$ in Eq.~\eqref{eq:Stot} over all the possible entries in $\mathcal{H}$. In particular, the optimal value of the net extracted work is given by application of Eq.~\eqref{eq:argmin} in  Eqs.~\eqref{eq:avwork} and \eqref{eq:averageEP}.

To simplify notation, below we will always assume an optimal choice of the lookup table index $n^*(\mathbf{m})$ and will sometimes write just $n$ or $n^\ast$ to refer to the optimal choice $n^*(\mathbf{m})$. The context should always make the precise meaning clear.

\subsection{Conditions for effective work extraction under control limitations}

Having specified the way to choose the different quenching Hamiltonians $H_n(\mathbf{x})$ in the set $\mathcal{H}$ for a given measurement outcome $\mathbf{m}$, we may implement feedback control to extract the average amount of work given in Eq.~\eqref{eq:avwork}. That implementation would make sense whenever a net work can be extracted from $\rho_0$ once the cost of information erasure is  discounted. The requirement for successful feedback control is hence $k_B T \langle S_\mathrm{tot} \rangle <  \Delta \mathcal{F}_H$, which can be conveniently written as:
\begin{equation} \label{eq:requirement}
    \langle F_{H_n}^{\rm eq} \rangle  - F_H^{\rm eq}  > \langle W_{\rm cost} \rangle + k_B T I_{X;M},
\end{equation}
where $\langle F_{H_n}^{\rm eq}\rangle = \sum_{n} p(n)  F_{H_n}^{\rm eq}$ is the average equilibrium free energy of the quenched Hamiltonian, and we identified the average work needed to implement the quenches $H \rightarrow H_n$ as
\begin{equation} \label{eq:wcost}
\langle W_{\rm cost} \rangle:= \sum_x \rho_0(\mathbf{x}) \left[\sum_n p_{{L}|\bold{x}}(n) H_n(\mathbf{x}) - H(\mathbf{x}) \right],    
\end{equation}
with $p_{{L}|\bold{x}}(n) := p(n | \mathbf{x}) = \sum_{\mathbf{m} \in \ell(n)} p_{M|\bold{x}}(\mathbf{m} | \mathbf{x})$ being the effect of the measurement noise on the choice of the partition $l(n)$ corresponding to the quenching Hamiltonian $H_n$.Equation~\eqref{eq:requirement} implies that a net work extraction is possible using feedback control if on average the (equilibrium) free energy difference between the quenching Hamiltonians and the original Hamiltonian $H$, overcomes the combined work cost of implementing the quench and posterior erasure of the information employed in the feedback control.

In many situations the work needed for implementing the quenches $\langle W_{\rm cost} \rangle$ in Eq.~\eqref{eq:wcost} would be non-zero, at least in the absence of perfect observation. For example, in a popular feedback strategy for discrete systems that was implemented in the single-electron box Maxwell's demon of Refs.~\cite{Koski2014,Koski2014P}, the quenches in the feedback loop are performed by significantly raising the energy of a level when it is not occupied, hence incurring in zero cost. However as soon as these measurements have some errors and the level might be occupied, there is a penalty cost in the quenching process that cannot be avoided~\cite{Koski2014,Koski2014P}. In more idealized situations such as the original Szilard engine, the work needed to make the quench is usually neglected $\langle W_\mathrm{cost} \rangle= 0$ in assuming that the control operations (there introducing partitions in the single-particle box) can be made, in principle, with no friction or other costs. Finally there might be situations in which that work is negative, i.e. work is extracted already (or only) during the quench.

\subsection{Work extraction from deterministic and random quenching operations}

Now consider a different scenario in which we try to extract work by using a either a pre-specified or a random quenching Hamiltonian $H_q \in \mathcal{H}$ which is actually independent of the measurement result $\mathbf{m}$. If the quenching Hamilronian is fixed to $H_q$, the average entropy production in Eq.~\eqref{eq:averageEP} simplifies to $\langle S_\mathrm{tot} \rangle = \sum_\mathbf{m} p(\mathbf{m}) D(\rho_{X|\mathbf{m}} || e^{- \beta H_q}/Z_q) = I_{X;M} + D(\rho_0 || e^{-\beta H_q})$, and hence the work that can be extracted in this case becomes:
\begin{align} \label{eq:detwork}
     {W}_q  &= \mathcal{F}_H(\rho_{0}) - {F}_H^{\rm eq} - k_B T D\left(\rho_0 \left\vert\left\vert \frac{e^{- \beta H_q}}{Z_q}\right.\right.\right) \nonumber \\ &= F_{H_q}^{\rm eq} - F_H^{\rm eq} - W_{\rm cost}^q,
\end{align}
and the whole expression becomes actually independent of $\mathbf{m}$. In the above equation, the last term stands for the work cost of the quench to $H_q$:
\begin{equation} \label{eq:qcost}
 W_{\rm cost}^q = \sum_x \rho_0(x) [H_q(\mathbf{x})- H(\mathbf{x})].
\end{equation}
Therefore the deterministic quenching can lead to an effective work extraction when the free energy difference between the quenching and original Hamiltonians is greater than the cost to quench, $F_{H_q}^{\rm eq} - F_H^{\rm eq} > W_{\rm cost}^q $.

We note that above, if we assume $H_q = H_0 := -k_B T \ln \rho_0$ (and hence $Z_q = 1$), the third therm in the first line of Eq.~\eqref{eq:detwork} vanishes and we can extract the full free energy contained in $\rho_0$, namely, $W_q = \mathcal{F}_H(\rho_{0}) - {F}_H^{\rm eq}$ in accordance to optimal work extraction without feedback control~\cite{Takara2010,Hasegawa2010,Esposito2011}. However, the Hamiltonians in the set $\mathcal{H}$ would be typically optimized to extract work from a specific measurement outcome ${m}$, and hence we assume in the following that $H_0 \notin \mathcal{H}$. 

Comparing the work extractable from imperfect feedback control (once the costs of information erasure has been subtracted) with ${W}_q$ we obtain the work gain due to using information in the quenching as:
\begin{align} \label{eq:feedbackgain} 
\langle \Delta W \rangle &:=  \langle {W} \rangle - k_B T I_{X;M} - {W}_q \nonumber \\ 
    &= \sum_{n\neq q} ~p(n)~F_{H_n}^{\rm eq} \\ 
         &~ + \sum_{\mathbf{x}}  \rho_0(\mathbf{x}) \left[H_q(\mathbf{x}) - \sum_n p_{{L}|\mathbf{x}}(n) H_{n}(\mathbf{x})\right]  \nonumber
\end{align}
where we used that $H_q \in \mathcal{H}$. The first sum of equilibrium free energies in the above expression is always positive. However, $\langle \Delta W \rangle$ can in principle take any sign depending on how $H_q$ and the set $\mathcal{H}$ is chosen, as well as on the level of noise in the measurement as captured by $p_{{L}|\bold{x}}(n)$. In particular, a net benefit in work extraction for the feedback control scenario requires that the energy after the quench $q$ is not very low compared with the average after-quenching energy. This will always be the case in the situation in which the work cost of quenching is negligible (that is, $\sum_\mathbf{x} \rho_0(\mathbf{x}) H \simeq \sum_\mathbf{x} \rho_0(\mathbf{x}) H_q$ for any $q$) and hence the second line in the above equation vanishes leading to $\langle \Delta W \rangle \geq 0$.

The expression in Eq.~\eqref{eq:feedbackgain} hence implies that using feedback control may allow the feedback controller not only to gain an extra amount of work $k_B T I_{X;M}$ related to the information acquisition from the measurements (which needs to be paid back at the end of the protocol), but it may also increase our ability to extract work from the original distribution $\rho_0$ (whenever the optimal quench $H_0:=-k_B T \ln \rho_0$ is not available), enhancing the amount of work we can obtain with respect to the case of just using a predefined quench in $\mathcal{H}$.

If we now assume that the Hamiltonian $H_q$ is sampled randomly from a distribution $r(q)$ independent from the measurement result, the average work extractable would be $\langle {W} \rangle_r = \sum_q r(q) W_q $ with $W_q$ in Eq.~\eqref{eq:detwork}. In that case the difference between the work extracted when choosing the correct quench and a random one from $r(q)$ is:
\begin{align} \label{eq:workgainr}
   \langle \Delta W \rangle_r &= \langle {W} \rangle - k_B T I_{X;M} - \sum_q r(q) {W}_q \nonumber \\ 
         &= \sum_{n} ~ \left[p(n) - r(n) \right]~F_{H_n}^{\rm eq} \\ 
         &+ \sum_{\mathbf{x}}  \rho_0(\mathbf{x}) \sum_n \left[ r(n) - p_{{L}| \mathbf{x}}(n) \right] H_{n}(\mathbf{x}).  \nonumber
\end{align}
We note that if we choose the random distribution above to mimic the marginal distribution to obtain the action from the measurement outcomes, $r(q) = p(q)$, the first term in the above equation vanishes. Moreover, since $p(n) = \sum_x p_{{L}| \mathbf{x}}(n) \rho_0(\mathbf{x})$ the second term is positive and we have a net gain $\langle \Delta W \rangle_p \geq 0$, remarking the convenience of using feedback control.

\subsection{The ``cheeseburger effect''}

\label{ssec:cheeseburger}

As a simple illustration of the value of information in feedback control scenarios, consider the following thought experiment. Suppose a cheeseburger is placed in either one of two boxes, with equal probability, so that a single bit of information is required to reveal the cheeseburger's location. If we only get one attempt to grab the cheeseburger, knowledge of which box it resides in doubles our energetic reward. So we  increase our free energy extraction by an amount that is approximately $10^{26}$ times greater than $k_BT\log 2$, just from learning a single bit. This differs greatly from typical considerations about the ``energetic value of a bit'', e.g., arising from the analysis of Szilard's engine. 

To understand this phenomenon, we first note that a cheeseburger's macrostate corresponds to a distribution of microstates representing the distinct arrangements of molecular ingredients composing that cheeseburger. We require two such distributions, corresponding to the two possible macrostates; we denote these distributions $\rho_A(\mathbf{x})$ for occupancy of the left box, and $\rho_B(\mathbf{x})$ for occupancy of the right box, with argument $\mathbf{x}$ representing a specific molecular arrangement. In the absence of information about in which box the cheeseburger is placed, its distribution would be ${\rho}_0(\bold{x})=[\rho_A(\mathbf{x})+\rho_B(\mathbf{x})]/2$. 

We now suppose that the macroscopic position of the cheesburger is accessible to measurement (e.g., by opening the boxes to see which one contains the cheeseburger), so that we identify the space $\mathbb{M}$ as consisting only of two possible discrete outcomes, left $m=A$ or right $m = B$, occurring with probabilities $p(A) = p(B) = 1/2$. Then by observing, say, $m=A$, the updated state of the system would be $\rho_{X|A}(\mathbf{x}) = \rho_A(\mathbf{x})$, and we may enact the ideal quenching Hamiltonian, $H_A(\mathbf{x})=-k_B T \log \rho_A(\mathbf{x})$. Similarly, if we obtain $m=B$, the state of the system gets updated to $\rho_{X|B}(\mathbf{x}) = \rho_B(\mathbf{x})$ and the quench $H_B(\mathbf{x})=-k_B T \log \rho_B(\mathbf{x})$ is performed. Notice that, unlike a measurement of the microstate (a distribution over $\mathbb{X}^{d_X}$), this is a measurement of the macrostate (a distribution over $\{A,B\}$). This is a similar measurement than in the case of Szilard's engine, where the measurement reveals which side of the box the particle is contained in but not the specific location and momentum of the particle.

Let's then assume that the cheeseburger is in one of the two possible (Bayesian posterior) states $\rho_{X|{m}}$, for $m= \{A, B\}$. If the outcome $m$ is accessible to the feedback controller a work $W(m)$ as defined in Eq.~\eqref{eq:ImperfectW} would be extracted, where in this case the optimal choice of quench in Eq.~\eqref{eq:argmin} is just $n^\ast(m) =m$. We assume that accessing the cheeseburger will allow for a positive work extraction, $W(m)>0$ for both $m=A,B$. The average work that the controller would be able to extract is then $\langle W \rangle = \sum_{m} p(m) W(m)> 0$. In contrast, if the controller has to randomly choose some action randomly, i.e. without access to the outcome $m$, the amount of extractable work will be reduced. In particular, let's first consider the situation in which even if the quenching Hamiltonian is chosen at random, $q = \{A, B\}$, with arbitrary probabilities $r(q)$, the true measurement outcome $m$ is going to be revealed to the experimenter just before action, so that the experimenter has the opportunity to halt the quench unless it is correct. Then when we average over measurement results, the work extracted is 
\begin{equation}
    \begin{aligned}        
\langle W^\prime \rangle_{r}&=\sum_{m}p(m)\sum_{q} r(q) W(m) ~\delta_{m, q}\\
&=\sum_{m} p(m) r(m) W(m) > 0.
    \end{aligned}
\end{equation}
The difference between the two situations is therefore 
\begin{equation}
\langle \Delta W^\prime \rangle_{r}=\sum_{{m}}p({m})W({m})\left[1-r(m)\right] > 0.
\end{equation}
This difference is strictly positive since $r(m) \leq 1$ and $\sum_m r(m) = 1$. It can be even macroscopically large if the initial system is far enough from thermal equilibrium, so that $W(m)$ is large on macroscopic scales for all $m$. That is, if the cheeseburger contains enough processable energy, typically $W(m) \simeq 300 {\rm cal}$ for a standard cheeseburger. 

An interesting variant of this analysis arises if we cannot halt an incorrect quench. In that case energetic penalties are incurred as a consequence of implementing a wrong quench, which may involve both the work cost of implementing the quench itself [see Eq.~\eqref{eq:qcost}] (the energy needed to pick up and eat the cheeseburger), and also the free energy lost in the subsequent relaxation to equilibrium (the putrefaction of the cheeseburger in the inaccessible box).

In this case, the average extracted work is
\begin{equation}
\langle {W} \rangle_{r} =\langle W^\prime \rangle_r - \sum_{m} p({m}) \sum_{q \neq m}r(q) W_{q}(m),
\end{equation}
where the second term in the first line represents the total work lost due to the implementation of the quench $H_k$ when the state of the system was $\rho_{X | {m}}$ (instead of $\rho_{X |k}$) and $W_q(m) = W(m) - k_B T D(\rho_{X|m} || \rho_{X|q})$ which in general can be either positive or negative.
 
The benefit of feedback control (i.e. observing and accurately quenching with the correct quench) over this strategy is given by
\begin{equation}
    \begin{aligned}
\langle \Delta{W} \rangle_r =& \langle \Delta W^\prime \rangle_r - k_B T I_{X;M} \\ &+ \sum_{m} p({m}) \sum_{q \neq m}r(q) W_{q}(m), 
    \end{aligned}
\end{equation}
which is consistent with Eq.~\eqref{eq:workgainr} for noise-free observation of the macroscopic variable $m = n$ and no limitations in the available quenches, i.e., $H_n= -\ln \rho_{X|n}$. Note that in the above expression we have incorporated the {\it erasure cost}, $k_B T I_{X;M} = k_B T [S(\rho_0) -S(\rho_{X|A})/2 - S(\rho_{X|B})/2]= k_B T S(p) = k_B T \log 2$. This is because, in contrast to the feedback control case, random quenching without observation requires no processing of information and hence a fair comparison requires taking that cost into account. The above equation tells us that despite having learned only one bit about $m$, the thermodynamic consequences for the distribution of the microstates $\mathbf{x}\in\mathbb{X}$ can be arbitrarily large.

\section{Optimal work extraction with control limitations}
\label{sec:optimal_extraction}
 
Importantly, rather than taking an arbitrary collection of Hamiltonians, in many situations we would be interested in constructing an optimal set of quenching Hamiltonians $\mathcal{H}$ with a limited dimension $N$, given an initial distribution $\rho_0(\mathbf{x})$ and given the conditional probability $p_{M| \mathbf{x}}(\bold{m})$ characterizing the noisy measurement of the system.

\begin{figure}
    \centering
    \includegraphics[scale=0.25, trim = 500 0 0 0,clip]{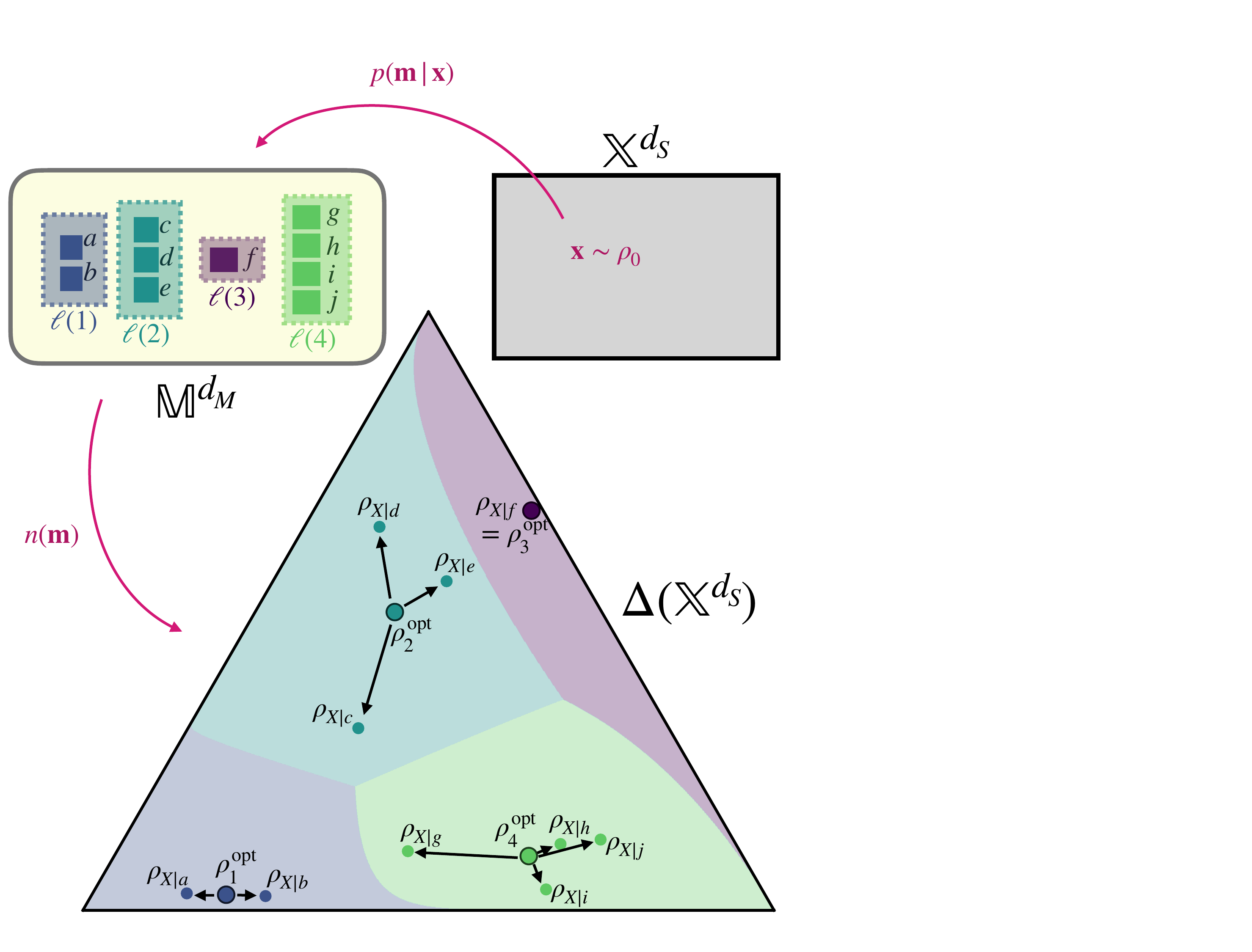}
    \caption{Illustration of the construction of an optimal quenching repertoire $\mathcal{H}$. We group all possible post-measurement distributions $\rho_{X|\mathbf{m}}$ in the simplex $\Delta(\mathbb{X}^{d_S})$ into $N$ sets according to a partition $\ell(n)$. The partition encodes a lookup table $n(\mathbf{m})$ from possible measurement results $\mathbf{m}$ (colored squares) in the space $\mathds{M}^{d_M}$ to the set of possible quenching actions. Measurement results $\mathbf{m}$ arise from the nonequilibrium system of interest (big grey square) by sampling from a noisy channel $p(\mathbf{m}|\mathbf{x})$, where system microstate $\mathbf{x}\in\mathbb{X}^{d_S}$ is distributed as $\rho_0$. The optimal quenches $\rho_n^{\rm opt}$ (large dots) within a given set $l(n)$ (green shaded area) is chosen by finding the distribution that minimizes the average entropy production (Kullback-Leibler divergence) when relaxing from all other post-measurement states within that set (small dots). Eq.~\ref{eq:optimalq2} is used to obtain $H_n^{\mathrm{opt}}$ from $\rho_n^{\mathrm{opt}}$. Shaded regions depict a Voronoi-like partitioning of $\Delta(\mathbb{X}^{d_S})$ according to KL divergence from $\{\rho_n\}_{n=1}^N$. In the illustration, $N=4$, and the partition is $\{a,b\},\{c,d,e\},\{f\},\{g,h,i,j\}$.}
    \label{fig:schematic} 
\end{figure}

As mentioned before, the measurement results $\mathbf{m}$ can be generically distributed into $N$ disjoint but otherwise arbitrary sets $l(n)$ with $n=1, ..., N$ to which one of the possible quenching Hamiltonians $H_n(\mathbf{x}) \in \mathcal{H}$ is attached by using the lookup table defined by the minimization in Eq.~\eqref{eq:argmin}. 
Constructing the optimal set $\mathcal{H}$ requires that the $H_n$ within each set $l(n)$ minimizes the expected entropy production across all such partition (see Fig.~\ref{fig:schematic}). The optimal form of $H_n(\mathbf{x})$ given a partition ${L}=\{\ell(n)\}_{n=1}^N$ is derived in Appendix \ref{app:optimal_quench_given_partition} and reads:
\begin{equation} \label{eq:optimalq}
    H_n(\mathbf{x})=- k_B T \ln \rho_n^{\rm opt}(\mathbf{x})
\end{equation}
where $\rho_n^{\rm opt}$ is a probabilistic mixture of post-measurement posterior distributions $\rho_{X|\mathbf{m}}$, as:
\begin{equation} \label{eq:optimalrho}
\rho_n^{\rm opt}(\mathbf{x}) := \sum_{\mathbf{m} \in l(n)} p(\mathbf{m}| n) \rho_{X|\mathbf{m}}(\mathbf{x}),
\end{equation}
with $p(\mathbf{m}| n) = \mathds{1}\{\mathbf{m} \in \ell(n)\} p(\mathbf{m})/p(n)$ the conditional probability of outcome $\mathbf{m}$ given knowledge of the set $n$ to which it pertains. Here $\mathds{1}\{\mathbf{m}\in \ell(n)\}$ is the indicator function for whether $\mathbf{m}$ resides within the partition block $l(n)$, and recall that $p(n)=\sum_{\mathbf{m} \in l(n)} p(\mathbf{m})$ is the probability that the measurement outcome $\mathbf{m}$ is inside the set $l(n)$. 
The distribution in Eq.~\eqref{eq:optimalrho} can be seen as mixed version of the post-measurement distribution $\rho_{X|\mathbf{m}}(\mathbf{x})$ after averaging over the detailed information about the outcome $\mathbf{m}$ when that information is unavailable but we only have information about within which set $n$ the outcome is.

Given the optimal form of the elements of the set $\mathcal{H}$ in Eq.~\eqref{eq:optimalq}, we can now compute the average entropy production (or work dissipated) in Eq.~\eqref{eq:averageEP} for a given choice of partition ${L}$:
    \begin{align} \label{eq:EPresult}
        \langle S_\mathrm{tot} \rangle &= \sum_\mathbf{m} p(\mathbf{m}) D(\rho_{X|\mathbf{m}}|| \rho_n^{\rm opt}) \\
       &= \sum_{\mathbf{m}} p(\mathbf{m})  \sum_{\mathbf{x}} \rho_{X|\mathbf{m}}(\mathbf{x}) \ln \frac{ \rho_{X|\mathbf{m}}(\mathbf{x})}{\rho_n^{\rm opt}(\mathbf{x})} \nonumber \\
        &= \sum_n p(n) \left(S(\rho_n^{\rm opt}) - \sum_{\mathbf{m} \in l(n)} p(\mathbf{m}|n) S(\rho_{X|\mathbf{m}})\right) \nonumber
    \end{align}
where we took the optimal index choice $n$ as the one verifying $\mathbf{m} \in \ell(n)$ for every outcome of the measurement. The above Eq.~\eqref{eq:EPresult} is one of our main results that relates the entropy production in the imperfect work extraction process to the increase in average Shannon entropy due to coarse-graining the information about the system state over the interval $\ell(n)$ for $n=1,..., N$.

 We note that in Eq.~\eqref{eq:EPresult}, the optimal quenching distribution $\rho_n^{\mathrm{opt}}(\bold{x})$ is a mixture of the post-measurement distributions $\rho_{X|\mathbf{m}}(\bold{x})$ across $\mathbf{m}$ weighted by $p(\mathbf{m}|n)$ [c.f. Eq.~\eqref{eq:optimalq}], and that the parenthesized term in Eq.~\eqref{eq:EPresult} is the mean entropy of $\rho_{X|\mathbf{m}}$ over the same mixing distribution $p(\mathbf{m}|n)$. As a consequence the entropy production can be rewritten as
 \begin{align} 
     \langle S_{\rm tot} \rangle = \sum_{n}p(n) J_{p(\mathbf{m}|n)}\left(\{\rho_{X|\mathbf{m}}\}_{\mathbf{m}\in\ell(n)}\right), \label{eq:JSD}
 \end{align}
with $J_{\mathbf{p}}(\mathbf{P}) := S(Q)-\sum_{i}p_iS(P_i) \geq 0$ denoting the generalized Jensen-Shannon divergence (JSD) among a weighted collection of distributions $\mathbf{P}=\{P_i\}_i$ with (normalized) weights $\mathbf{p}=\{p_i\}_i$~\cite{Lin91}. Here $Q(\bold{x})=\sum_{i}p_iP_i(\bold{x})$ denotes the mixture distribution of $\{P_i(\bold{x})\}_i$ with weights $\{p_i\}_i$. (Note, $\sum_i p_i=1$ and $\sum_\bold{x} P_i(\bold{x})=1$.) In Eq.~\eqref{eq:JSD} we identified $P_i(\bold{x}) = \rho_{X| \bold{m}}(\bold{x})$ and $p_i = p(\bold{m}|n)$  with $i = \bold{m} \in l(n)$, leading to $Q(\bold{x}) = \rho_n^{\rm opt}(\bold{x})$, c.f. Eq~\eqref{eq:EPresult}. The Jensen-Shannon divergence and its generalizations have been employed in a variety of fields ranging from bioinformatics~\cite{Sims09} to the social sciences~\cite{DeDeo14} and more recently in machine learning~\cite{Hossein24}. The above result in Eq.~\eqref{eq:JSD} implies that the optimal partition is one that minimizes the generalized JSD among the distributions grouped into a common partition block $\ell(n)$, averaged over the action $n$ taken.

A second, illuminating way of rewriting $\langle S_\mathrm{tot} \rangle$ in Eq.~\eqref{eq:EPresult} is in terms of informational quantities. A key fact is 
that the distribution $\rho_n^{\rm opt}(\bold{x})$ in Eq.~\eqref{eq:optimalrho} is the marginal probability of finding the system in state $\bold{x}$ given that the measurement result is inside the partition $l(n)$. More precisely, we can rewrite the entropy production as the conditional mutual information~\cite{CoverThomasBook} between the system state and the measurement result, given the 
choice of quenching Hamiltonian according to the partitions $l(n)$:

\begin{align} \label{eq:mutualcond}
    \langle S_\mathrm{tot} \rangle &= \sum_n p(n) \sum_\bold{x} \sum_{\bold{m} \in l(n)} p(\bold{m}|n) \rho_{X|\bold{m}}(\bold{x}) \ln \left( \frac{\rho_{X|\bold{m}}(\bold{x})}{\rho_n^{\rm opt}(\bold{x})} \right) \nonumber \\
    &= \sum_n p(n) \sum_\bold{x} \sum_{\bold{m} \in l(n)} P(\bold{x}, \bold{m} | n) \ln \left( \frac{P(\bold{x}, \bold{m} | n)}{\rho_n^{\rm opt}(\bold{x}) p(\bold{m} | n)} \right) \nonumber \\
    &= I_{X;M | {L}} \geq 0,
\end{align}
where in the second line we expanded the joint conditional probability $P( \bold{x} , \bold{m} | n) = p(\bold{m} | n) \rho_{X|\bold{m}}(\bold{x})$, and inside the logarithm we multiplied and divided by $p(\bold{m}|n)$.

 Eq.~\eqref{eq:mutualcond} says that achieving exactly zero entropy production requires that for all the partitions $l(n)$, there must be zero conditional mutual information $I_{X;M | {L}}$ between $\bold{x}$ and $\bold{m}$ given that the measurement result $\mathbf{m}$ is in the set $l(n)$. In other words, it requires the post-measurement distribution $\rho_{X|\bold{m}}(\bold{x})$ to actually be independent of the measurement result $\bold{m}$ within every partition $l(n) \in {L}$. Note that this effect arises \textit{only} because of the restriction on the number of possible actions. Without such a restriction, this information-theoretic loss of the maximal amount of free energy the the agent can extract would not exist.

 Notice though that such a condition cannot be satisfied without spoiling the possibility of performing feedback control, which in general requires correlations between $\bold{m}$ and $\bold{x}$, c.f. Eq.~\eqref{eq:avwork}. In other words, zero entropy production requires that the measurements provide no other information than the one necessary to deduce the actual set $n$ to which the measurement belongs. In that case the distributions $\rho_{X|\bold{m}}(\bold{x})$ only depends on $n$, which would imply actually $\rho_n^{\rm opt}(\bold{x}) = \rho_{X|\bold{m}}(\bold{x})$. That condition would be (trivially) satisfied when the number of possible quenching Hamiltonians coincide with the number of measurement results, $N = |\mathbb{M}^{d_M}|$, as expected.
 
By plugging Eq.~\eqref{eq:mutualcond} into Eq.~\eqref{eq:avwork} we obtain the average extractable work:
\begin{align} \label{eq:avworkfinal}
    \langle {W} \rangle =&~ k_B T \left[I_{X;M} - I_{X;M | {L}} \right] 
    + \Delta \mathcal{F}_H.
\end{align}
 The term $I_{X;M} - I_{X;M | {L}}$ is sometimes called the ``interaction information". In general, it can be either positive or negative~\cite{CoverThomasBook}. Therefore, an efficient measurement generating high correlations between $\bold{m}$ and $\bold{x}$ in general improves work extraction, in that it provides more information $I_{X;M}$ and in the limit of perfect correlation allows the extraction of all the free energy in the initial state. At the same time, spurious correlations as characterized by the conditional mutual information $I_{X;M|{L}}$ may spoil the ability to extract work by increasing the entropy production. 
 
Taking into account the erasure of information after the feedback control protocol implies, nonetheless, that the optimal choice of partition corresponds to the one minimizing $I_{X;M | {L}}$. In other words, it corresponds to one in which the measurements generate the minimum mutual information between the measurement outcomes and the system state within the partitions $l(n)$, hence allowing the correct choice of the quenching Hamiltonians $H_n (\bold{x}) = - k_B T \ln \rho_n^{\rm opt}(\bold{x})$ but avoiding, as much as possible, the correlations that can not be used in the feedback control operation.
 
Finally, we can straightforwardly obtain an upper bound on the entropy production as
\begin{equation}
\sum_n p(n) S(\rho_n^{\rm opt}) \geq \langle S_\mathrm{tot} \rangle,
\end{equation}
which follows from the concavity of entropy, $S(\rho_n^{\rm opt}) \geq \sum_{\bold{m} \in l(n)} p(\bold{m}|n) S(\rho_{X|\bold{m}}) \geq 0$. The above upper bound establishes respectively a lower bound on the average work that can be extracted, c.f. Eq.~\eqref{eq:avwork}.

\section{Some applications}
\label{sec:applications}

In the following we describe various situations that may arise in different contexts fitting within the general framework introduced above. In these applications the work extraction procedure always uses feedback control, but the kind of measurements employed and the limitations vary.

\subsection{Feedback control with memory limitations and noise-free measurements}
\label{ssec:perfect_observation}

As a standard scenario of general interest in stochastic thermodynamics we consider in the following the case in which the measurement results $m_i \in \mathbf{m}$ are associated one to one to the system variables $x_i \in \mathbf{x}$ and hence $\mathbb{M} = \mathbb{X}$ and $d_X=d_M$. Moreover, even if we are allowed to optimally choose the set of quenching Hamiltonians in the set $\mathcal{H}$, we assume operation with a finite memory where the number of quenches $N$ is generically much smaller than the state space of the system $N < |\mathbb{X}^{d_X}|$, and hence likewise smaller than the possible measurement results $|\mathbb{M}^{d_M}|=|\mathbb{X}^{d_X}|$.

In this context, we here assume the case of ideal (perfect) measurements on the system where $p(\mathbf{m}) = \rho_0(\mathbf{x})$ and $\rho_{X|\mathbf{m}}(\mathbf{x}) = \delta_{\mathbf{m},\mathbf{x}} =\prod_{i= 1}^d \delta_{m_i, x_i}$. The optimal quench set $\mathcal{H}$, according to Eqs.~\eqref{eq:optimalq} and \eqref{eq:optimalrho} then reduces to: 
\begin{equation} \label{eq:optimalq2}
    H_n(\mathbf{x})=- k_B T \ln \left(\frac{\rho_0 (\mathbf{x})}{p^{\rm cg}_0(n)}  \mathds{1}\{\mathbf{x}\in \ell(n)\} \right),
\end{equation}
where $\mathds{1}\{\mathbf{x}\in \ell(n)\}$ is the indicator function within the partition $l(n)$, and $p^{\rm cg}_0(n) := \sum_{x \in \ell(n)} \rho_0(\mathbf{x})$ is the coarse-grained version of $\rho_0(\mathbf{x})$ according to partition $\ell(n)$. That is, the above quenching Hamiltonian takes finite values in the interval $\mathbf{x} \in l(n)$ and it is infinite otherwise. Here above the optimal post-quench distribution $\rho_n^{\rm opt}(\mathbf{x})$ is just the expression inside the parenthesis.

In this case of ideal measurements, a more intuitive result can be obtained from Eq.~\eqref{eq:EPresult} (see Appendix.~\ref{app:perfect_observation}), namely,
\begin{align} \label{eq:entropy-drop}
 \langle S_\mathrm{tot} \rangle &= \sum_\mathbf{m} \rho_0(\mathbf{m}) D(\delta_{\mathbf{m}, \mathbf{x}} || e^{-\beta H_{n}}) \nonumber \\ 
 &= \sum_n p^{\rm cg}_0(n) S(\rho_n^{\rm opt}) = S(\rho_0) - S(p_0^{\rm cg}),
\end{align}
which is just the drop in Shannon entropy when coarse-graining the original start distribution $\rho_0(\mathbf{x})$ over the partitions $\ell(n)$. The above equation indicates that the extra dissipation (work lost) in this case is entirely due to the loose of resolution in the initial distribution $\rho_0$ due to the coarse-graining imposed by the partition $l(n)$ over the finite memory of the controller. 

Noting that $p_0^{\rm cg}(n)$ is a probability distribution with $N$ distinct outcomes, we have $S(p_0^{\mathrm{\mathrm{cg}}}) \le \log N$ which is attained if and only if the partition $\ell(n)$ is such that $p_0^{\rm cg}$ is uniformly distributed. Therefore, we obtain the bound:
\begin{equation} \label{eq:idealbound}
     \langle S_\mathrm{tot} \rangle \ge S(\rho_0) - \log N,
\end{equation}
across all possible partitions $\ell(n)$ with fixed dimension $N$. As a result we see that the optimal design for the distribution of measurement outcomes consists of a partition $l(n)$ that is probabilistically equitable with respect to the initial distribution $\rho_0$. That is
\begin{equation}
    p_0{^{\rm cg}}(n) = \sum_{x \in \ell(n)} \rho_0(\mathbf{x}) = \sum_{x \in \ell(n')} \rho_0(\mathbf{x}) = p_0{^{\rm cg}}(n')
\end{equation}
for all $n,n'$ in the set, in which case the bound in Eq.~\eqref{eq:idealbound} is attained. Additionally, we notice it may be possible to further tighten the bound above by examination of the optimal choice of partition given $\rho_0$.

\begin{figure}[t]
    \centering
    \includegraphics[width=0.95\linewidth]{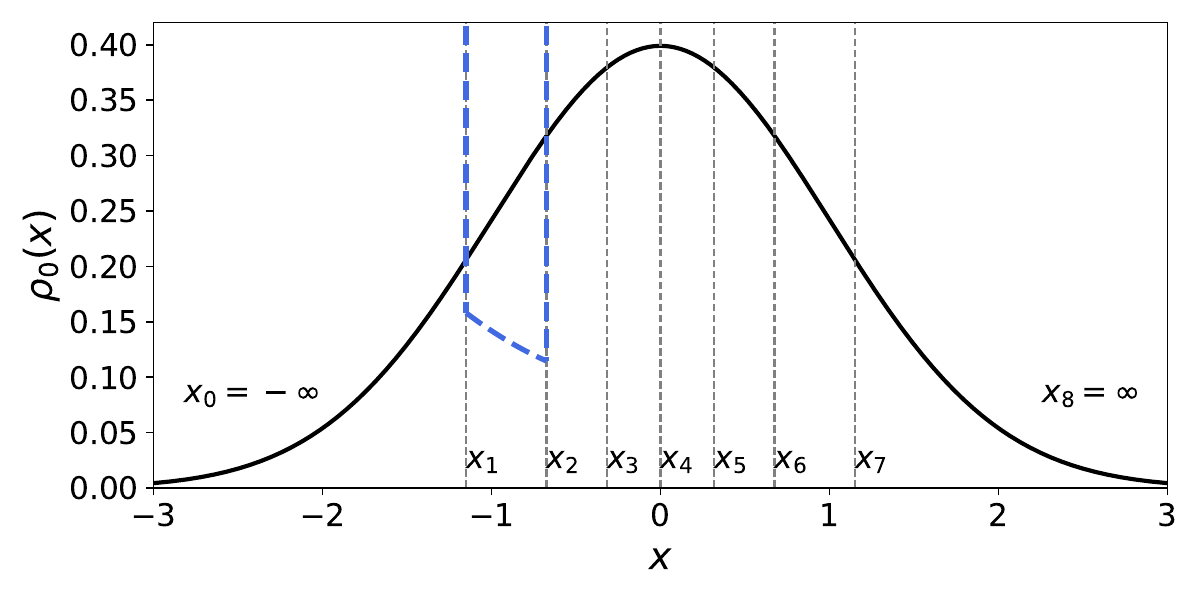}
    \caption{Optimal partition $\ell(n)$ for a particle in one dimension using $N=8$ contiguous regions, $(x_0,x_1]$, $(x_1,x_2]$, ..., $(x_{7},x_8)$ according to Eq.~\eqref{eq:1Dxn}. The solid line corresponds to the initial distribution, here taken to be $\rho_0(\mathbf{x})= e^{-x^2/2}/\sqrt{2\pi}$ while the blue dashed curve corresponds to the 
    quenching Hamiltonian $H_n(\mathbf{x})$ in Eq.~\eqref{eq:optimalq2} for $n=2$ (not normalized, for visibility), equalling $\infty$ outside of $(x_1,x_2]$.}
    \label{fig:partition}
\end{figure}

As a simple example of interest, let us consider a Brownian particle in one dimension described by its position $x$ in $\mathbb{R}$ with initial density $\rho_0(x)$. This situation has been often implemented in the laboratory using colloidal particles trapped by laser beams in the context of stochastic thermodynamics~\cite{Toyabe2010,Berut2012experimental,Roldan2014,ciliberto2017}.
We assume perfect observation, so that $p(m|x)=\delta(m-x)$, leading to the $n$th optimal quenching Hamiltonian in the set $\mathcal{H}$ as given in Eq.~\eqref{eq:optimalq2}, with $\rho_0 ^{\rm cg} = \int_{\ell(n)}\rho_0(x')dx'$ and $\{\ell(n)\}$ a partition of $\mathbb{R}$. For simplicity, we now make the restriction that $\ell(n)$ is a {\it contiguous} partition, i.e., that each partition block has the form $\ell(n)=(x_{n-1},x_n]$ for some $x_0<x_1<...<x_N$, with $N$ size of the Hamiltonian repertoire, and with the last partition block having the form $\ell(N)=(x_{N-1},x_N)$ rather than $(x_{N-1},x_N]$. We take $x_{0}=-\infty$ and $x_N=\infty$ so that $\cup_{n}\ell(n)=\mathbb{R}$; if we took any $x_0>-\infty$ or $x_N<\infty$, our quenching repertoire would have smaller support than $\rho_0$, yielding unnecessary dissipation.

In this case, the unique way to achieve optimal quenching is to set partition $\ell(n)$ as $N$ intervals in $\mathds{R}$ with boundaries $x_k$ for $k=0, 1, 2, ..., N$ at
\begin{equation}
\label{eq:1Dxn}
x_k=F_0^{-1}\left(\frac{n}{N}\right),
\end{equation}
with $F_0(x)=\int_{-\infty}^{x}\rho_0(x')dx'$ denoting the cumulative probability of $X<x$ under $\rho_0$ (see Appendix~\ref{app:perfect_observation} for a proof).

The optimal partition $\ell(n)$ is shown in Fig.~\ref{fig:partition} for $N=8$ possible quenches. The optimal quenches inherit the form of the initial distribution $\rho_0$ within each interval as illustrated with the blue dashed line in the second partition ($n=2$).

\subsection{Work extraction from random initial distributions}
\label{ssec:random_distributions}

A second related but different scenario of interest is the case in which the initial distribution $\rho_0(\mathbf{x})$ of the system is not completely known to the feedback controller. 

Uncertainty in the initial distribution amounts to a {\it random macrostate}, from which the randomness of the microstates given the macrostate is subsidiary. In this setting, the measurement is {\it about} the distribution (i.e., information regarding which of several possible distributions may describe the system), rather than {\it of} the distribution (i.e., a noisy measurement of the system microstate).

In particular, we assume that the initial distribution $\rho_0^{(\alpha)}(\bold{x})$ is sampled from a finite set labeled by $\alpha$ according to some prefixed probability $p_\alpha$, with $\sum_\alpha p_\alpha = 1$. In this case we consider that the feedback controller performs some (imperfect) measurement of the system concerning the value of $\alpha$ of the original distribution, rather than the state $\mathbf{x}$. Then the conditional probability capturing the effect of external noise is of the form ${p}(m|\alpha)$, where now the result of the measurement $m$ is a single number related to the values of the parameter $\alpha$, occurring with (marginal) probability $p(m) = \sum_\alpha p(m | \alpha) p_\alpha$. Such a situation is to be compared with the cases in which no information about the parameter $\alpha$ is available, implying an effective mixture 
\begin{equation}
\label{eq:mixture}
    \bar{\rho}_0 = \sum_\alpha p_\alpha \rho_0^{(\alpha)},
\end{equation}
and should also be compared to the case in which perfect knowledge of the initial distribution $\rho_0^{(\alpha)}$ is assumed.

After performing the noisy measurement of $\alpha$, the controller can gain knowledge of the initial distribution of the system. The conditional probability of obtaining a possible $\alpha$ value given the measurement result $m$ is ${p}(\alpha|m)=p_\alpha {p}(m|\alpha)/p(n)$, from which the updated system distribution ${\rho}_{X|m}(x)$ can be computed as:
\begin{equation}
\label{eq:tilde_rho_e_random_beta}
    {\rho}_{X|m}(\mathbf{x}) = \sum_{\alpha } {p}(\alpha|m) \rho_0^{(\alpha)}(\mathbf{x}).
\end{equation}
Notice that the controller obtains perfect knowledge of the initial distribution if the measurement is noise-free, i.e. when ${p}(m|\alpha)=\mathds{1}\{m=\alpha\}$ leading to $\rho_{X|m}(\mathbf{x}) = \rho_0^{(\alpha(m))}$ with $\alpha(m)$ denoting the parameter unequivocally related to the measurement outcome. On the other hand, 
zero knowledge from the measurement is obtained if ${p}(m|\alpha)$ is independent of $\alpha$, so that we obtain $\rho_{X|m}(\mathbf{x})= \bar{\rho}_{0}(\mathbf{x})$. 

Assuming no further limitations in the memory or action of the controller, the repertoire of Hamiltonians defining the quench set $\mathcal{H} = \{H_m(\mathbf{x})\}$ to optimally extract work from the initial state is given by the updated states after measurement, that is, $H_m(\mathbf{x}) = - \ln \rho_{X|m}(\mathbf{x})$. As a consequence of the mismatch between that distribution and the ``true'' initial distribution, entropy production will be generated during the feedback control protocol also in this case. Such entropy production is as before related to the relaxation from the (generally unknown) distribution $\rho_0^{(\alpha)}$ to the equilibrium state $e^{-\beta H_m(\mathbf{x})}$ after the quench. Following our previous arguments, the average entropy production due to the uncertainty in the parameter $\alpha$ is therefore given by
\begin{equation}
\langle S_\mathrm{tot} \rangle = \sum_{\alpha} p_\alpha \sum_{m} {p}(m|\alpha)\mathrm{D}(\rho_0^{(\alpha)} || {\rho}_{X|m})\ge 0,
\end{equation}
with equality if and only if the channel ${p}(m|\alpha)$ is noise-free, that is, if ${\rho}_{X|m}(\mathbf{x}) = \rho_0^{(\alpha(m))}(\mathbf{x})$. On the other hand such entropy production is maximized in the zero-knowledge case, leading to $\langle S_\mathrm{tot} \rangle = \sum_\alpha p_\alpha D(\rho_0^{(\alpha)} || \bar{\rho}_0)$.

We now consider the combination of the situations of random initial distributions as described above and control limitations. Since each measurement outcome $m$ is unambiguously related to an optimal post-quench equilibrium distribution $\rho_{X|m}(\mathbf{x})$, our previous logic will apply. If there are $|\mathds{M}|$ possible measurements $m\in \{1,..., |\mathds{M}|\}$, and $N$ possible quenching Hamiltonians, $n\in\{1,...,N\}$, where $N<|\mathds{M}|$, then as before we need to design a repertoire and lookup table $n(m)$ mapping from the measurement results to the associated quenches in $\mathcal{H}$. Given a partition ${L} = \{ \ell(n)\}_{n=1}^N$, similar calculations to those of Appendix~\ref{app:optimal_quench_given_partition}  apply to obtain an optimal quenching Hamiltonian $H_n(\mathbf{x}) = -k_B T \ln \rho_n^{\rm opt}(\mathbf{x})$ with:
\begin{equation}
\label{eq:rho_n_opt_alpha}
\rho_n^{\mathrm{opt}}(\mathbf{x})=\sum_{\alpha}p(\alpha|n)\rho^{(\alpha)}_0(\mathbf{x}),
\end{equation}
with $p(\alpha|n)=p_\alpha p(n|\alpha)/p(n)$, where $p(n)=\sum_\alpha p(n|\alpha)$, and $p(n|\alpha)=\sum_{m\in\ell(n)}P(m|\alpha)$. Details are reported in Appendix~\ref{app:random_initial}, where we also show the equivalence of Eq.~\eqref{eq:rho_n_opt_alpha} above and Eq.~\eqref{eq:optimalrho}, with use of the measurement-based expression Eq.~\eqref{eq:tilde_rho_e_random_beta}. That is, we show that we may equivalently express the form of the optimal post-quench equilibrium $\rho^{\mathrm{opt}}_n$ as a probabilistic mixture of $\rho_{X|\mathbf{m}}$ across $\mathbf{m}$ (given $n$) or as a probabilistic mixture of $\rho_0^{(\alpha)}$ across $\alpha$ (given $n$).

Then likewise the average entropy production under control limitations reads:
\begin{equation}
\label{eq:stot_random_initial}
 \langle S_\mathrm{tot} \rangle = \sum_\alpha p_\alpha \sum_{m} p(m|\alpha) D(\rho_0^{(\alpha)} || \rho_{n^*(m)}^{\rm opt}),
 \end{equation}
which also exhibits a representation in terms of JSD and conditional mutual information akin to Eqs.~\eqref{eq:JSD} and \eqref{eq:mutualcond} respectively.

In particular, under the optimal action, $\langle S_{\mathrm{tot}}\rangle$ can be expressed as the average of a weighted multi-distributional Jensen-Shannon divergence, namely
\begin{equation}
\label{eq:JSD_expression_alpha}
\langle S_{\mathrm{tot}}\rangle=\sum_{n}p(n)J_{p_{\alpha|n}}(\{\rho_0^{(\alpha)}\}_{\alpha}).
\end{equation}

\begin{figure}
    \centering
    \includegraphics[scale=0.33,trim = 600 20 200 0,clip]{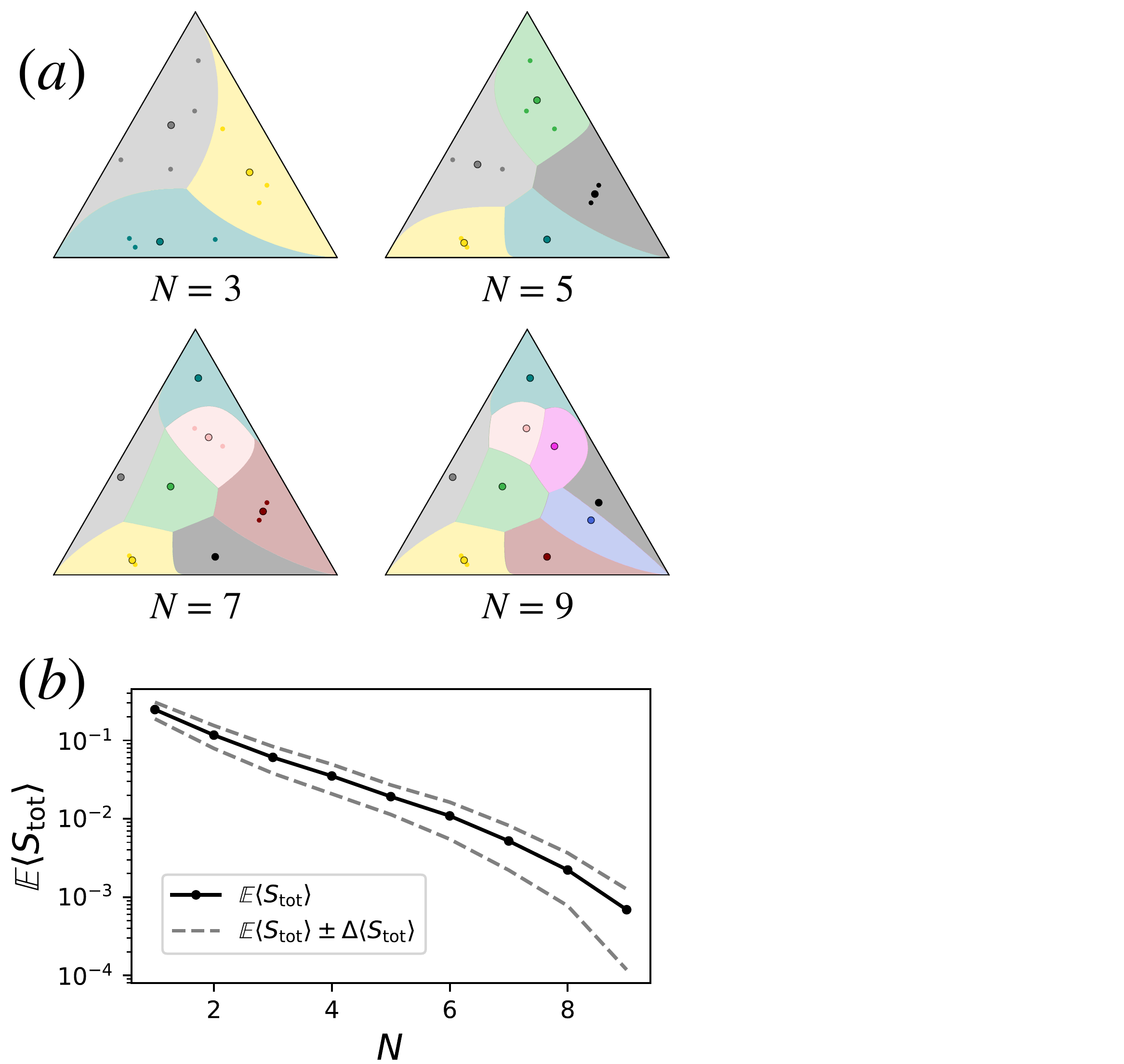}
    
    \caption{(a) Optimal partitions and post-quench equilibria for $N=3,5,7,9$ actions and $10$ possible initial distributions. The distributions $\{\rho_0^{(\alpha)}\}_{\alpha=1}^{10}$ are depicted as small dots, colored according to their assigned partition; large dots of corresponding colors represent the optimal quenching choices $\rho_n^{\mathrm{opt}}$. Additionally, regions are shaded based on the minimization of the KL divergence between $\{\rho_n^{\mathrm{opt}}\}_{n=1}^N$ and the points therein, akin to a Voronoi diagram \cite{onishi1997voronoi,nielsen2020voronoi}. (b) Expected dissipation under optimal partition as a function of the number of actions $N$ when the round-truth distributions $\{\rho_0^{(\alpha)}\}_{\alpha=1}^{10}$ are sprinkled randomly into the unit $3$-simplex. $\mathbb{E}\langle S_{\mathrm{tot}}\rangle$ and $\Delta\langle S_{\mathrm{tot}}\rangle$ denote the mean and standard deviation of $\langle S_{\mathrm{tot}}\rangle$ in Eq.~\eqref{eq:stot_random_initial}.}
    \label{fig:partitioning}
\end{figure}

The JSD in Eq.~\eqref{eq:JSD_expression_alpha} at any given $n$ is the dissipation associated with action $n$ in response to the noisy measurement of $\alpha$. Accordingly, optimal partitions will cluster nearby distributions in the probability simplex (see Fig.~\ref{fig:partitioning} for an illustrative example).

Additionally, akin to Eq.~\eqref{eq:mutualcond}, we can re-express $\langle S_{\mathrm{tot}}\rangle$ in terms of a conditional mutual information between the system micro and macro states for a given a partition, $\langle S_{\mathrm{tot}}\rangle=I_{X,A;L}$, with $X,A,L$ denoting the random variables whose values are denoted $\mathbf{x},\alpha,n$ (see Appendix~\ref{app:random_initial} for details). We note that here the macrostates $\alpha$ play the role of the measurement results $\mathbf{m}$ in expression~\eqref{eq:mutualcond}. Therefore, we can conclude that minimum dissipation in this scenario requires a minimum amount of extra correlations between microstates and macrostates, other than those strictly needed to perform the choice of the optimal action.

We demonstrate this scenario with a numerical example, obtaining the optimal partitions for a set of possible initial nonequilibrium distributions indexed by $\alpha=1,...,10$. We denote these distributions as $\{\rho^{(\alpha)}_0\}_{\alpha=1}^{10}$, with each $\rho_0^{(\alpha)}$ being a $3$-outcome distribution (for visualizability in the unit simplex). Moreover, we consider equal probabilities for each of them, $p_\alpha=1/10$ for all $\alpha$. We then numerically scan across all distinct partitionings of the $10$ distributions (macrostates) into $N$ response actions, computing the dissipation under the optimal choice in Eq.~\eqref{eq:rho_n_opt_alpha}.

In Fig.~\ref{fig:partitioning}a we depict the optimal partition for different number of allowed actions (memory size limitations) $N=3,5,7,9$. The different colored areas visually represent which quenching Hamiltonian will be chosen for the different distributions $\rho_0^{(\alpha)}$ (represented as small dots). The optimal quenching distributions for each partition, Eq.~\eqref{eq:rho_n_opt_alpha}, are represented by the larger dots. In Fig.~\ref{fig:partitioning}b we visualize the dissipation resulting from the optimal action repertoire with a number of available actions $N<10$. We show the mean $\mathds{E}[\langle S_\mathrm{tot} \rangle]$ and standard deviation $\Delta[\langle S_\mathrm{tot} \rangle]$ of the entropy production in the feedback control process across numerous samples of the choice of initial distributions. As can be observed, when $N$ grows from $1$ to $9$, the dissipation decreases confirming the optimal choice for the partitions.

\section{Conclusions and Discussion}
\label{sec:conclusions}

In this work we have developed a thermodynamic framework to describe the interplay of measurement and feedback control operation in systems with restricted capabilities preventing completely optimal free energy extraction.

We formulated this problem by considering a nonequilibrium distribution that can be noisily measured, in response to which one of a limited number of actions can be taken, in the form of {\it quenches}, determining when extraction of a net amount of work remains possible. In particular, the problem boils down to the minimization of the expected Kullback-Leibler divergence between the would-be optimal response and the actual response resorted to; this requires a partition of scenarios into response actions, and in addition, a specific form of those response actions given the partition. We derived an explicit expression for the latter, finding that the optimal post-quench equilibrium is a probabilistic mixture of would-be optimal quenches across the scenarios grouped together into a common response. Furthermore, we obtained an expression for the dissipated work in the case of optimal quenching, finding that it equals the average across possible actions of the multi-distributional Jensen-Shannon divergence among the distributions whose mixture composes the optimal action. This reduced the general problem of optimal response construction to a simpler and more specific search problem across partitions of scenarios into actions.

Additionally, we have considered a variety of different situations of restricted and permitted measurements and actions, including the employment of random quenching strategies or deterministic quenching strategies. We obtained the expected work extraction in these situations and derived expressions comparing free energy extraction capacity across several such scenarios, discussing the favorability of different strategies in different situations. That allowed us to explore and explain the fact that a single bit’s thermodynamic worth can far exceed $k_B T \log 2$, exemplified by a cheeseburger residing uncertainly in one of two locations — when learning that one bit, we may gain half a cheeseburger’s worth of calories in comparison to the expectation in the case of choosing the location at random.

The type of control limitations considered here are arguably ubiquitous in experimental settings feedback control in nanosystems, where a complete control of the system of interest is generally not possible and the intrinsic stochastic behavior due to the impact of thermal fluctuations make difficult to have perfect measurements~\cite{Bechoefer2024}. In such situations the type of configurations from which work could be optimally or even efficiently extracted is reduced~\cite{Maillet2019,Lucero2021,Saha2022} and many experiments exploring the thermodynamics of small systems typically restrict themselves to the case of initial equilibrium states, that is, as in Maxwell's demon setup. The theory developed here should help to provide a precise description of work extraction from arbitrary nonequilibirum initial states under noisy observation and lack of perfect control.

Our findings have relevance to many real-world systems, including physical, biological, and social in which action limitations strongly impact the free energy extraction process. Indeed these systems doesn't need to be necessarily small~\cite{Parrondo2001}. For instance, numerous biological systems efficiently harvest free energy under stochastic environmental conditions with constraints on possible actions. For example, individual organisms with a finite number of innate behaviors, e.g., a bacterium or an insect, must selectively apply these behaviors under a multitude of conditions \cite{jiang2022neural,barron2015decision}. Evolutionary pressures hence arguably favor the development of effective action repertoires, both in terms of the actions themselves and in terms of the mapping of scenarios to actions. Housing a perfect innate response for all situations would be exorbitantly costly; instead, it is advantageous to develop multi-purpose behaviors \cite{bell1990searching}.

In addition, such limitations are abundant in social systems. For example, many modern hunter-gatherer groups maintain social norms that help facilitate collective behavior, including the deterrence of free-riders and the encouragement of demand sharing (i.e., sharing food according to need; \cite{lewis2014high}). Thus, when an individual hunter-gatherer finds food or is successful in a hunt, their energy extraction actions are limited: they might eat some of the food immediately, but must save the rest to share with their camp. Further, while most hunter-gatherer populations store or preserve food, the extent of this storage is typically limited \cite{armit1992hunter, hamilton2024food}. In contrast, individuals living in farming societies rely on being able to freely choose among multiple actions. Namely, to consume food immediately, or store it for later use. The increasingly complex technology available to farming and urbanized societies \cite{hamilton2024institutional} suggests that having access to a larger action repertoire may be crucial to explaining the massive increase in per-person energy use between the ancient and modern world \cite{krausmann2016transitions}.

There are many rich ways to extend the analysis in this paper, to capture even more
features of real world systems that try to extract free energy from their environment.
One obvious extension is to consider the case where there is more than just one agent who is extracting the free energy. In this extension the different agents would run quench-then-relax protocols on distributions involving different physical variables in the environment. It might also be that each agent observes some features of importance to some other ones, and so would need to communicate their information to the other agents, incurring a thermodynamic cost. Another avenue of future research arises from the fact that
in this paper considering only a single agent we do \textit{not} model the thermodynamic cost of the physical system that computes what action to take based on a given measurement. In particular we would need to consider such issues in a system that is used more than once, so that some of the variables in the agent need to be reinitialized at the beginning of each run.

\ 

\section*{Acknowledgements}
H.\,H., D.\,H.\,W., A.\,J.\,S., and C.\,K. acknowledge support from the Santa~Fe Institute. D.~W. acknowledges support from US NSF Grant CCF-2221345. G.~ M. acknowledges support from the ``Ramón y Cajal'' program (No. RYC2021-031121-I), the CoQuSy project (No. PID2022-140506NB-C21), and the Mar\'ia de Maeztu project (No. CEX2021-001164-M) funded by MCIU/AEI/10.13039/501100011033 and European Union NextGenerationEU/PRTR.

\appendix

\section{Optimal set of quenching Hamiltonians for a given partition}
\label{app:optimal_quench_given_partition}
Here we derive what the optimal quenching Hamiltonians are for a given lookup table encoded by partition of observations into actions. In so doing we reduce the general problem of repertoire construction to the simpler (yet still nontrivial) problem of finding an optimal partition.

Recall, the post-measurement would-be optimal quench (under no limitations) is the Bayesian posterior $\rho_{X|\mathbf{m}}(\mathbf{x})=\rho_0(\mathbf{x}) P(\mathbf{m}|\mathbf{x})/p(\mathbf{m})$, with $p(\mathbf{m})=\sum_{\mathbf{x}}\rho_0(\mathbf{x})P(\mathbf{m}|\mathbf{x})$. The observation $\mathbf{m}$ is grouped into the action indexed by $n^*(\mathbf{m})$. The inverse of this mapping defines a partition $\ell(n)$ into blocks indexed by $n$. In response to any observation $\mathbf{m}$ in a given partition block $\ell(n)$, a common response is employed, namely, the quenching Hamiltonian $H_n$, or equivalently, the associated equilibrium distribution $\rho_n^{\mathrm{opt}}=e^{-\beta H_n}/Z_n$.

The mean dissipated work $\langle S_{\mathrm{tot}}\rangle$ is the mean Kullback-Leibler divergence between the would-be optimal quench $\rho_{X|\mathbf{m}}$ and the actual quench resorted to, $\rho^{\mathrm{opt}}_{n^*(\mathbf{m})}$.  That is,
\begin{equation}
\begin{aligned}
\langle S_{\mathrm{tot}}\rangle &= \sum_{\mathbf{m}}p(\mathbf{m})D(\rho_{X|\mathbf{m}}||\rho_{n^*(\mathbf{m})}^{\mathrm{opt}})\\
&=\sum_{n}p(n)\sum_{\mathbf{m}\in \ell(n)}p(\mathbf{m}|n)\mathrm{D}(\rho_{X|\mathbf{m}}||\rho_n^{\mathrm{opt}}),
\end{aligned}
\end{equation}
with $p(n)=\sum_{\mathbf{m}\in\ell(n)}p(\mathbf{m})$ being the coarse-grained probability of performing action $n$, i.e., the probability that $\mathbf{m}\in\ell(n)$. We seek the repertoire $\mathcal{H}=\{H_n\}_n$ corresponding to the collection of distributions $\{\rho_{n}^{\mathrm{opt}}\}_{n}$ minimizing $\langle S_{\mathrm{tot}}\rangle$, where we take the partition as fixed.

Note,
\begin{equation}
\frac{\mathrm{d}\langle S_{\mathrm{tot}}\rangle}{\mathrm{d}\rho_n^{\mathrm{opt}}(\mathbf{x})}=p(n)\sum_{\mathbf{m}\in\ell(n)}p(\mathbf{m}|n)\frac{\mathrm{d}D(\rho_{X|\mathbf{m}}||\rho_n^{\mathrm{opt}})}{\mathrm{d}\rho_n^{\mathrm{opt}}(\mathbf{x})},
\end{equation}
with
\begin{equation}
\begin{aligned}
\frac{\mathrm{d}D(\rho_{X|\mathbf{m}}||\rho_n^{\mathrm{opt}})}{\mathrm{d}\rho_n^{\mathrm{opt}}(\mathbf{x})}=-\frac{\rho_{X|\mathbf{m}}(\mathbf{x})}{\rho_n^{\mathrm{opt}}(\mathbf{x})}.
\end{aligned}
\end{equation}
Therefore,
\begin{equation}
\begin{aligned}
    \frac{\mathrm{d}\langle S_{\mathrm{tot}}\rangle}{\mathrm{d}\rho_n^{\mathrm{opt}}(\mathbf{x})}&=-p(n)\sum_{\mathbf{m}\in\ell(n)}p(\mathbf{m}|n)\frac{\rho_{X|\mathbf{m}}(\mathbf{x})}{\rho_n^{\mathrm{opt}}(\mathbf{x})}\\
&=-\frac{\sum_{\mathbf{m}\in\ell(n)}p(\mathbf{m})\rho_{X|\mathbf{m}}(\mathbf{x})}{\rho_n^{\mathrm{opt}}(\mathbf{x})}.
\end{aligned}
\end{equation}
Introducing a set of Lagrange multipliers $\{\lambda_n\}_{n}$, we consider Lagrangian
\begin{equation}
L=\langle S_{\mathrm{tot}}\rangle+\sum_{n}\lambda_n \sum_{\mathbf{x}}\rho_n^{\mathrm{opt}}(\mathbf{x}),
\end{equation}
with the latter terms included to enforce that all $\{\rho_n^{\mathrm{opt}}\}$ are normalized. Minimizing with respect to any given one of the variables $\{\rho_n^*(\mathbf{x})\}_{n,\mathbf{x}}$, we have
\begin{equation}
  \begin{aligned}
      0&=\frac{\partial L}{\partial \rho_n^{\mathrm{opt}}(\mathbf{x})}\\
      &=-\frac{\sum_{\mathbf{m}\in\ell(n)}p(\mathbf{m})\rho_{X|\mathbf{m}}(\mathbf{x})}{\rho^{\mathrm{opt}}_{n}}+\lambda_n,\\
  \end{aligned}  
\end{equation}
from which we have
\begin{equation}
    \rho_n^{\mathrm{opt}}(\mathbf{x})=\frac{1}{\lambda_n}\sum_{\mathbf{m}\in\ell(n)}p(\mathbf{m})\rho_{X|\mathbf{m}}(\mathbf{x}).
\end{equation}
The values of $\{\lambda_n\}_n$ are determined by normalization:
\begin{equation}
    \begin{aligned}
    \lambda_n&=\sum_{\mathbf{x}}\sum_{\mathbf{m}\in\ell(n)}p(\mathbf{m})\rho_{X|\mathbf{m}}(\mathbf{x})\\
&=\sum_{\mathbf{m}\in\ell(n)}p(\mathbf{m})\\
&=p(n)\\
    \end{aligned}
\end{equation}
and thus
\begin{equation}
\begin{aligned}
    \rho_n^{\mathrm{opt}}(\mathbf{x})&=\frac{1}{p(n)}\sum_{\mathbf{m}\in\ell(n)}p(\mathbf{m})\rho_{X|\mathbf{m}}(x)\\
    &=\sum_{\mathbf{m}\in\ell(n)}p(\mathbf{m}|n)\rho_{X|\mathbf{m}}(\mathbf{x}).
\end{aligned}
\end{equation}
Thus $\rho_n^{\mathrm{opt}}(\mathbf{x})$ is a probabilistic mixture of the would-be optimal responses $\rho_{X|\mathbf{m}}(\mathbf{x})$, with mixture distribution being that of $\mathbf{m}$ conditional on action $n$ being taken. For the associated quenching Hamiltonian, $H_n(\mathbf{x})=-k_BT\log \rho_n^{\mathrm{opt}}(\mathbf{x})$. Therefore, the problem of finding an optimal repertoire is reduced to the problem of finding an optimal partition.

\section{The case of perfect observation}
\label{app:perfect_observation}

Herein we consider the case of perfect observation, in which case $\rho_{X|\mathbf{m}}(\mathbf{x})=\delta_{\mathbf{m},\mathbf{x}}$. The distribution of measurements is therefore
\begin{equation}
    \begin{aligned}
    p(\mathbf{m})&=\sum_{\mathbf{x}}\delta_{\mathbf{x},\mathbf{m}}\rho_0(\mathbf{x})\\
    &=\rho_0(\mathbf{m}),
    \end{aligned}
\end{equation}
with $\mathbf{m}$ a valid argument of $\rho_0$ because $\mathbb{M}^{d_M}=\mathbb{X}^{d_S}$ to accommodate perfect observation. In this case the distribution representing the $n$th partition block is given by
\begin{equation}
\begin{aligned}
\rho_n^{\mathrm{opt}}(\mathbf{x})&=\sum_{\mathbf{m}\in\ell(n)}p(\mathbf{m}|n)\delta_{\mathbf{m},\mathbf{x}}\\
    &=p(\mathbf{x}|n).
\end{aligned}
\end{equation}
The conditional distribution of measurements given that they lead to action $n$ is proportional to $\rho_0$, windowed to support $\ell(n)$. Namely,
\begin{equation}
p(\mathbf{x}|n)=\frac{\mathds{1}\{\mathbf{x}\in\ell(n)\}\rho_0(\mathbf{x})}{p^{\mathrm{cg}}(n)},
\end{equation}
with $p^{\rm cg}_0(n) := \sum_{x \in \ell(n)} \rho_0(\mathbf{x})$ being the coarse-grained version of $\rho_0(\mathbf{x})$. Note that the partition block $\ell(n)$ of possible measurements $\mathbf{m}\in\ell(n)$ is now also a partition of possible environmental states $\mathbf{x}\in\ell(n)$, since $\mathbb{M}^{d_M}=\mathbb{X}^{d_S}$ under perfect observation.

The mean dissipated work is given by
\begin{equation}
\langle S_{\mathrm{tot}}\rangle=\sum_{n}p^{\rm cg}(n)J_{p(\mathbf{m}|n)}(\{\rho_{X|\mathbf{m}}\}),
\end{equation}
with, in this case,
\begin{equation}
    \begin{aligned}
        J_{p(\mathbf{m}|n)}(\{\rho_{X|\mathbf{m}}\})&=S(\rho_n^{\mathrm{opt}})-\sum_{\mathbf{m}\in\ell(n)}p(\mathbf{m}|n)S(\delta_{\mathbf{x},\mathbf{m}})\\
        &=-\sum_{\mathbf{x}\in\ell(n)}\frac{\rho_0(\mathbf{x})}{p^{\rm cg}(n)}\log \frac{\rho_0(\mathbf{x})}{p^{\rm cg}(n)}.
    \end{aligned}
\end{equation}
Therefore,
\begin{equation}
\begin{aligned}
\langle S_{\mathrm{tot}}\rangle&=\sum_{n}p(n)\left(-\sum_{\mathbf{x}\in\ell(n)}\frac{\rho_0(\mathbf{x})}{p^{\rm cg}(n)}\log \frac{\rho_0(\mathbf{x})}{p^{\rm cg}(n)}\right)\\
&=-\sum_{n}\sum_{\mathbf{x}\in\ell(n)}\rho_0(\mathbf{x})\log \rho_0(\mathbf{x})+\sum_{n}p^{\rm cg}(n)\log p^{\rm cg}(n)\\
&=S(\rho_0)-S(p^{\rm cg}).
\end{aligned}
\end{equation}

Noting that $p^{\rm cg}$ is a probability distribution with $N$ distinct possible outcomes, we have
\begin{equation}
    S(p^{\rm cg})\le \log N,
\end{equation}
with $\log N$ achieved if and only if the partition $\ell(y)$ is such that $p^{\rm cg}$ is uniformly distributed across the set of $N$ possible actions. We call such a partition (probabilistically) {\it equitable}. Therefore, we arrive at the bound
\begin{equation}
     \langle S_{\mathrm{tot}}\rangle \ge S(\rho_0)-\log N,
\end{equation}
At any given finite $N$ and for a given distribution $\rho_0$, partitions achieving this bound may or may not exist.

\section{Details on the particle in $\mathbb{R}$ example}
\label{app:1D}

Herein we show that in the setting explored in the example of Sec.~\ref{ssec:perfect_observation} (1D, perfect observation, contiguous partition, $\rho_0$ has full support on $\mathbb{R}$), the unique way to achieve optimal quenching is to set partition boundaries at the positions given in Eq.~\eqref{eq:1Dxn}.

 Recall that an optimal partition must be probabilistically equitable, so that the distribution over measurements is maximally entropic: $S(p(m))=\log |\mathbb{M}|=\log N$. That is, we want the total probability of landing in $\ell(n)$ to be $1/N$, for each $n\in\{1,...,N\}$. 

By the fact that $\rho_0(x)>0$ for all $x\in \mathbb{R}$, we have $\frac{d}{dx}F_0(x)>0$ for all $x\in\mathbb{R}$. Therefore $F_0(x)$ is monotonically increasing and uniquely invertible; we denote $F_0^{-1}:[0,1]\rightarrow\mathbb{R}$ as the inverse of $F_0$. Recall, $x_0=-\infty$, as we argued that our quenching Hamiltonian repertoire $\mathcal{H}$ should (collectively) reproduce the support of $\rho_0$. The location of $x_1$ is thus uniquely determined by the condition that $\mathbb{P}(X<x_1)=1/N$. That is,
\begin{equation}\begin{aligned}
\label{eq:x1}
\int_{-\infty}^{x_1}\rho_0(x)dx&=F_0(x_1)=\frac{1}{N}\\
\Rightarrow x_1 = &F_0^{-1}\left(\frac{1}{N}\right).
\end{aligned}\end{equation}
Now by induction we can obtain $x_2,...,x_N$. Suppose the inductive hypothesis that for some $n\in\{2,...,N\}$ we have $x_{n-1}=F_0^{-1}(\frac{n-1}{N})$. Then, by monotonicity of $F_0$, and the optimality requirement of $\mathbb{P}(X\in \ell(n))=1/N$, the uniquely determined position of $x_n$ will satisfy
\begin{equation}
\label{eq:xn}
\int_{x_{n-1}}^{x_n}\rho_0(x)dx=F_0(x_n)-F_0(x_{n-1})=\frac{1}{N} .
\end{equation}
By hypothesis, $x_{n-1}=F_0^{-1}(\frac{n-1}{N})$, and thus $F_0(x_{n-1})=\frac{n-1}{N}$. Therefore, our requirement on $x_n$ (Eq.~\ref{eq:xn}) becomes
\begin{equation}\begin{aligned}
F_0(x_n)-\frac{n-1}{N}&=\frac{1}{N} \\
\Rightarrow F_0(x_n)&=\frac{n}{N},
\end{aligned}\end{equation}
from which by solving for $x_n$ we obtain Eq.~\ref{eq:1Dxn}. Since we showed earlier that $x_1=F_0^{-1}(1/N)$ (in Eq.~\ref{eq:x1}), we have by induction that Eq.~\ref{eq:1Dxn} holds for all $n\in\{2,...,N\}$. (Including the case of $n=N$ in which we recover the already assumed $x_N=\infty$ at $F_0^{-1}(1)$, since $F(\infty)=1$.) See Fig.~\ref{fig:partition} in the main text for an example of the optimal partition boundary locations in the case where $\rho_0(x)$ is a standard normal density.

We stress that the uniqueness of this solution is lost if (a) the full support requirement fails, in which case regions of $\rho_0(x)=0$ may be present, corresponding to regions of $F_0(x)=\mathrm{const}$ (in which boundaries may be freely moved, retaining a partition identical in function). Uniqueness is also lost if (b) the contiguity requirement is dropped. We conjecture that this degeneracy vanishes in generic cases of noisy measurement, in which case localized noise will favor partitions with smaller boundaries.

\begin{table*}[h]
\begin{center}
\begin{tabular}{ |p{2cm}||p{15cm}|}
 \hline
 \multicolumn{2}{|c|}{Table of notation} \\
 \hline
 \centering Symbol & Definition \\
 \hline
  \centering $\mathcal{H}$& Set of quenching Hamiltonians $\{H_k\}_{k=1}^N$ for feedback control\\
 \centering ${L}$& Set of partitions $\{l(n)\}_{n=1}^N$ grouping the measurement outcomes \\
 \centering $\rho_0(\bold{x})$   &  Initial marginal distribution of the system states $\bold{x}$  \\
  \centering $\rho_H^{\rm eq}(\mathbf{x})$&  Thermal equilibrium state for the system states $\bold{x}$ with Hamiltonian $H(\bold{x})$  \\
 \centering $\rho_{X|\bold{m}}(\bold{x})$& Conditional distribution of the system states $\bold{x}$ given measurement result $\bold{m}$  \\
 \centering $\rho_n^{\rm opt}(\bold{x})$& Distributions leading to optimal quenching Hamiltonians $H_n= -k_B T \log \rho_n^{\rm opt}$\\
 \centering $p(\bold{m})$ & Marginal distribution of the measurement outcomes $\bold{m}$ \\
 \centering $p_{M|\bold{x}}(\bold{m})$    & Conditional distribution of measurement outcomes $\bold{m}$ for given system state $\bold{x}$ \\
 \centering $n^\ast(\bold{m})$& Optimal choice for the quenching Hamiltonian given a measurement result $\mathcal{m}$\\
 \centering $p(n)$& Marginal distribution for obtaining partition $l(n)$ in the measurement \\
 \centering $p_{{L}|\bold{x}}(n)$& Conditional probability distribution for obtaining partition $l(n)$ given that the system is in state $\bold{x}$\\
 \centering $\rho_0^{\rm cg}(n)$& Marginal distribution for obtaining partition $l(n)$ in the noise-free scenario \\
\centering $p_\alpha$& Probability of component $\alpha$ in the uncertain initial distributions scenario\\
 \centering $\rho_0^{(\alpha)}(\bold{x})$& Component $\alpha$ of the initial marginal distribution in the uncertain initial distributions scenario \\
  \centering $\bar{\rho}_0(\bold{x})$& Effective mixture of marginal distributions in the uncertain initial distributions scenario \\
 \centering $S(\rho)$& Shannon entropy of distribution $\rho(\bold{x})$ \\
 \centering $D(\rho || \sigma)$& Kullback-Leibler divergence (relative entropy) between distributions $\rho(\bold{x})$ and $\sigma(\bold{x})$ \\
 \centering $J_{\mathbf{p}}(\mathbf{P})$& Generalized Jensen-Shannon divergence among a weighted collection of distributions $\mathbf{P}$ with weights $\mathbf{p}$ \\
 \centering $I_{X;M}$& Mutual information between the system states and measurement results \\
 \centering $I_{X;M|{L}}$& Conditional mutual information between the system and measurement results given a set of partitions ${L}$\\
 \centering $F_H^{\rm eq}$& Equilibrium free energy of the system for Hamiltonian $H(\bold{x})$  \\
 \centering $\mathcal{F}_H(\rho)$& Nonequilibrium free energy of state $\rho$ for Hamiltonian $H(\bold{x})$  \\
 \centering $\Delta \mathcal{F}_H$& Available nonequilibrium free energy in the initial state, $\mathcal{F}_H(\rho_0) - F_H^{\rm eq}$ \\
   \centering $S_\mathrm{tot}(\bold{m})$& Entropy production (free energy loss) in the feedback control process for measurement outcome $\bold{m}$ \\
 \centering $\langle S_\mathrm{tot} \rangle$& Average entropy production (average free energy loss) in the feedback control process after many runs \\
 \centering $W_{\rm ideal}(\mathbf{m})$& Work extracted in the ideal feedback control protocol for measurement outcome $\bold{m}$\\
 \centering $\langle W_{\rm ideal}\rangle$& Average work extracted in the ideal feedback control protocol after many runs\\
 \centering $W(\mathbf{m})$& Work extracted in the general feedback control protocol with limitations for measurement outcome $\bold{m}$ \\
 \centering $\langle W \rangle$& Average work extracted in the general feedback control protocol after many runs \\
\centering $W_{\rm cost}^q$& Work cost for performing a predefined quench $H_q(\bold{x})$ \\
 \centering $\langle W_{\rm cost} \rangle$& Average work cost for performing the quenches after many runs \\
 \centering $\langle \Delta W \rangle$& Average work gain from feedback control with respect to a deterministic quench after many runs \\
 \centering $\langle \Delta W \rangle_r$& Average work gain from feedback control with respect to random quenching using distribution $r(q)$ \\ 
 \hline
\end{tabular}
\end{center}
\caption{Table of notation for the main symbols used in the paper.} \label{TABLE}
\end{table*}

\section{Additional calculations for an arbitrary random initial distribution}
\label{app:random_initial}

Herein we provide details on the derivation of results for the scenario of feedback action based on a noisy measurement of a distribution-specifying parameter $\alpha$, considered in Section~\ref{ssec:random_distributions}. The parameter $\alpha$, sampled with probability $p_{\alpha}$, specifies the initial nonequilibrium distribution $\rho_0^{(\alpha)}$. Then a noisy measurement $m$ of $\alpha$ is produced with probability $p(m|\alpha)$, after which the feedback action $n=n^\star(m)$ is employed in response. Here again a lookup table $n^{\star}$ determines a partition $\{\ell(n)\}_n$ by $m\in\ell(n)\Rightarrow n^\star(m)=n$. 

First we note that the total entropy production in Eq.~\eqref{eq:stot_random_initial} can be re-expressed as follows
\begin{equation}
\begin{aligned}
\label{eq:Stot_alpha_1}
\langle S_{\mathrm{tot}}\rangle&=\sum_{\alpha}p_{\alpha}\sum_{m}P(m|\alpha)D(\rho^{(\alpha)}_0||\rho_{n(m)}^{\mathrm{opt}})\\
&=\sum_{\alpha}p_\alpha\sum_{n}p(n|\alpha)D(\rho^{(\alpha)}_0||\rho_n),
\end{aligned}
\end{equation}
with $p(n|\alpha)=\sum_{m\in\ell(n)}P(m|\alpha)$. Let us then define the conditional probability that parameter $\alpha$ was sampled given that action $n$ is taken
\begin{equation}
\begin{aligned}
p_{\alpha|n}=\frac{p(n|\alpha)p_\alpha}{p(n)},
\end{aligned}
\end{equation}
with $p(n)=\sum_{\alpha}p_\alpha p(n|\alpha)$ the marginal probability of taking action $n$. Then continuing from Eq.~\eqref{eq:Stot_alpha_1} we obtain
\begin{equation}
\begin{aligned}
\label{eq:Stot_alpha_2}
\langle S_{\mathrm{tot}}\rangle&=\sum_{\alpha,n}p_\alpha p(n|\alpha)D(\rho^{(\alpha)}_0||\rho_n^{\mathrm{opt}})\\
&=\sum_{n}p(n)\sum_{\alpha}p_{\alpha|n}D(\rho^{(\alpha)}_0||\rho_n^{\mathrm{opt}}).
\end{aligned}
\end{equation}

The optimal post-quench equilibrium $\rho_n^{\mathrm{opt}}$ minimizes the summand $\sum_{\alpha}p_{\alpha|n}D(\rho^{(\alpha)}_0||\rho_n^{\mathrm{opt}})$ in the above Eq.~\eqref{eq:Stot_alpha_2}, and thus by the derivation of Appendix~\ref{app:optimal_quench_given_partition} we have that the solution has a probabilistic mixture form:
\begin{equation}
\begin{aligned}
\label{eq:rho_n_opt_alpha_app}
\rho_n^{\mathrm{opt}}=\sum_{\alpha}p_{\alpha|n}\rho^{(\alpha)}_0.
\end{aligned}
\end{equation}
which was reported as Eq.~\eqref{eq:rho_n_opt_alpha} in the main text. Note the difference between Eq.~\eqref{eq:optimalrho} and Eq.~\eqref{eq:rho_n_opt_alpha_app} above; we will reconcile these below [see Eq.~\eqref{eq:last_step}].

In the optimal scenario of Eq.~\eqref{eq:rho_n_opt_alpha}, the dissipated work becomes
\begin{equation}
\begin{aligned}
\label{eq:Stot_JSD_alpha}
\langle S_{\mathrm{tot}}\rangle&=\sum_{n}p(n)\sum_{\alpha}p_{\alpha|n}D\left(\rho^{(\alpha)}_0\left\vert\left\vert \sum_{\alpha}p_{\alpha|n}\rho^{(\alpha)}_0\right.\right.\right)\\
&=\sum_{n}p(n)\sum_{\alpha}p_{\alpha|n}\sum_{x}\rho_0^{(\alpha)}(\mathbf{x})\log\frac{\rho_0^{(\alpha)}(\mathbf{x})}{\rho_n^{\mathrm{opt}}(\mathbf{x})}\\
&=\sum_np(n)\left(S(\rho_n^{\mathrm{opt}})-\sum_\alpha p_{\alpha|n}S(\rho_0^{(\alpha)})\right)\\
&=\sum_np(n)J_{p_{\alpha|n}}(\{\rho_0^{(\alpha)}\}_\alpha).
\end{aligned}
\end{equation}
The latter expresses that the dissipation is the average weighted multi-distributional JSD among the possible ground truth distributions $\{\rho_0^{(\alpha)}\}_\alpha$ as weighted by the posterior probability of $\alpha$ given $n$.

Additionally, we can re-express $\langle S_{\mathrm{tot}}\rangle$ in terms of a conditional mutual information. The conditional mutual information among three random variables $X,Y,Z$ is given by the general expression
\begin{equation}
\begin{aligned}
I_{X,Y;Z}=\sum_{z} P(z) \sum_{x,y}P(x,y|z)\log\frac{P(x,y|z)}{P(x|z)P(y|z)}.
\end{aligned}
\end{equation}
We will show below that $\langle S_{\mathrm{tot}}\rangle=I_{X, A;\mathcal{L}}$, with shorthand $X,A,\mathcal{L}$ for the random variables whose values are being denoted $\mathbf{x},\alpha, n$. First we define the joint distribution of $(\alpha,\mathbf{x})$ given $n$, denoted 
\begin{equation}
\begin{aligned}
p(\alpha,\mathbf{x}|n)&=p_{\alpha|n}\rho_0^{(\alpha)}(\mathbf{x}).\\
\end{aligned}
\end{equation}
The conditional distribution of $\mathbf{x}$ given $n$ is
\begin{equation}
\begin{aligned}
\sum_{\alpha}p(\alpha,\mathbf{x}|n)&=\sum_{\alpha}p_{\alpha|n}\rho_0^{(\alpha)}(\mathbf{x})=\rho_{n}^{\mathrm{opt}}(\mathbf{x}).\\
\end{aligned}
\end{equation}
The overall probability of action $n$ is 
\begin{equation}
\begin{aligned}
p(n)&=\sum_{\alpha,m}p_\alpha P(m|\alpha)\delta_{n,n^\star(m)}\\
&=\sum_{\alpha}p_\alpha\sum_{m\in\ell(n)}P(m|\alpha)\\
&=\sum_{\alpha}p_\alpha p(n|m),
\end{aligned}
\end{equation}
that is, it equals the optimal distribution for the quenching Hamiltonian in Eq.~\eqref{eq:rho_n_opt_alpha_app}. Therefore, we have
\begin{equation}
\begin{aligned}
I_{X, A;L}&=\sum_n p(n) \sum_{\alpha,\mathbf{x}}p(\alpha,\mathbf{x}|n)\log\frac{p(\alpha,\mathbf{x}|n)}{p_{\alpha|n}\rho_{n}^{\mathrm{opt}}(\mathbf{x})}\\
&=\sum_np(n)\left(\sum_{\alpha,\mathbf{x}}p_{\alpha|n}\rho_0^{(\alpha)}(\mathbf{x})\log\frac{\rho_0^{(\alpha)}(\mathbf{x})}{\sum_{\alpha}p_{\alpha|n}\rho_0^{(\alpha)}(\mathbf{x})}\right)\\
&=\sum_np(n)\left(S(\rho_n^{\mathrm{opt}})-\sum_{\alpha}p_\alpha S(\rho_0^{(\alpha)})\right),\\
\end{aligned}
\end{equation}
reproducing Eq.~\eqref{eq:Stot_JSD_alpha}. Thus the mean dissipated work is the conditional mutual information between the post-quench equilibrium state $\mathbf{x}$ and the ground truth index $\alpha$, with conditionality on the action taken $n$.

As a consistency check, let's reconcile Eq.~\eqref{eq:rho_n_opt_alpha} with Eq.~\eqref{eq:optimalrho} of the manuscript, to demonstrate the compatibility of the measurement-based expressions with the case of arbitrary initial distributions. Given the measurement of $m$, we have a posterior over $\alpha$ of the form
\begin{equation}
\begin{aligned}
p(\alpha|m)&=\frac{P(m|\alpha)p_\alpha}{p(m)}\\
&=\frac{P(m|\alpha)p_\alpha}{\sum_{\alpha}P(m|\alpha)p_\alpha},
\end{aligned}
\end{equation}
from which we can calculate the post-measurement distribution as
\begin{equation}
\begin{aligned}
\rho_{X|m}(\mathbf{x})&=\sum_{\alpha} p(\alpha|m)\rho_0^{(\alpha)}(\mathbf{x}).
\end{aligned}
\end{equation}
Thus by following the reasoning of Appendix~\ref{app:optimal_quench_given_partition}, the $n$th optimal action should result in a post-quench equilibrium of the form:
\begin{equation}
\begin{aligned}
\label{eq:measurement_expression}
\rho_n^{\mathrm{opt}}(\mathbf{x})&=\sum_{m}p(m|n)\rho_{X|m}(\mathbf{x}).
\end{aligned}
\end{equation}
Eq.~\eqref{eq:measurement_expression} is of the form of a mixture distribution across measurement outcomes, as opposed to the expression derived above, Eq.~\eqref{eq:rho_n_opt_alpha_app}, which is a mixture distribution across the hidden parameter $\alpha$. We next show the equivalence of the two. 

Rewriting Eq.~\eqref{eq:measurement_expression}, 
\begin{equation} \label{eq:measurement_expression_2}
\begin{aligned}
\rho_n^{\mathrm{opt}}(\mathbf{x})&=\sum_{m}p(m|n)\rho_{X|m}(\mathbf{x})\\
&=\sum_{m}\frac{\mathds{1}\{m\in\ell(n)\}p(m)}{p(n)}\rho_{X|m}(\mathbf{x})\\
&=\frac{\sum_{m\in\ell(n)}p(m)\rho_{X|m}(\mathbf{x})}{\sum_{m\in\ell(n)}p(m)}.
\end{aligned}
\end{equation}
Now noting that $p(m)=\sum_{\alpha}P(m|\alpha)p_\alpha$, we have
\begin{equation}
\begin{aligned}
\sum_{m\in\ell(n)}p(m)&=\sum_{m\in\ell(n)}\sum_{\alpha}P(m|\alpha)p_\alpha\\
&=\sum_{\alpha}\left(\sum_{m\in\ell(n)}P(m|\alpha)\right)p_\alpha\\
&=\sum_{\alpha}p(n|\alpha)p_\alpha =p(n).
\end{aligned}
\end{equation}
Similarly, we also have
\begin{equation}
\begin{aligned}
\sum_{m\in\ell(n)}p(m)\rho_{X|m}(\mathbf{x}) &= \sum_{m\in\ell(n)}p(m)\left(\sum_{\alpha} p(\alpha|m)\rho_0^{(\alpha)}(\mathbf{x})\right)\\
&=\sum_{m\in\ell(n)}\sum_{\alpha}P(m|\alpha)p_\alpha\rho_0^{(\alpha)}(\mathbf{x})\\
&=\sum_{\alpha}p_{\alpha}\rho_0^{(\alpha)}(\mathbf{x})\sum_{m\in\ell(n)}P(m|\alpha)\\
&=\sum_{\alpha}\rho_0^{(\alpha)}(\mathbf{x})p(n|\alpha)p_{\alpha},
\end{aligned}
\end{equation}
Therefore, replacing the above two expressions on Eq.~\eqref{eq:measurement_expression_2} we obtain:
\begin{equation}
\begin{aligned}
\label{eq:last_step}
\rho_n^{\mathrm{opt}}(\mathbf{x})&=\sum_{m}p(m|n)\rho_{X|m}(\mathbf{x})\\
&=\frac{\sum_{\alpha}\rho_0^{(\alpha)}(\mathbf{x})p(n|\alpha)p_{\alpha}}{p(n)}\\
&=\sum_{\alpha}\rho_0^{(\alpha)}(\mathbf{x})p(\alpha|n),
\end{aligned}
\end{equation}
with $p(\alpha|n)=p(n|\alpha)p_\alpha/p(n)$.

\subsection{Relation to original measurement scheme}
\label{app:orig_from_random_initial}

Finally, we show how the original measurement scenario considered in Sec.~\ref{sec:framework} is equivalent to the random initial distributions scenario when $\alpha \sim p_\alpha$ is assumed to be a random microstate $\mathbf{x}\sim \rho_0$, again passed through a noisy channel [$P(\mathbf{m}|\mathbf{x})$ rather than $P(m|\alpha)$], and with the associated set of ground truth distributions $\{\rho_0^{(\alpha)}\}_\alpha$ being set equal to $\{\delta_{\mathbf{x},\mathbf{x}'}\}_{\mathbf{x}}$.

Under these circumstances ($\alpha\rightarrow \mathbf{x},p_\alpha\rightarrow \rho_0(\mathbf{x}), \rho_0^{(\alpha)}(\mathbf{x})\rightarrow \rho_0^{(\mathbf{x})}(\mathbf{x}')=\delta_{\mathbf{x},\mathbf{x}'}$), we have
\begin{equation}
\begin{aligned}
p(\alpha|m)\rightarrow p(\mathbf{x}|\mathbf{m})&=\frac{P(\mathbf{m}|\mathbf{x})\rho_0(\mathbf{x})}{\sum_{\mathbf{x}}P(\mathbf{m}|\mathbf{x})\rho_0(\mathbf{x})}\\
&=\frac{P(\mathbf{m}|\mathbf{x})\rho_0(\mathbf{x})}{p(\mathbf{m})},
\end{aligned}
\end{equation}
and thus the posterior is
\begin{equation}
\begin{aligned}
\rho_{X|\mathbf{m}}(\mathbf{x}')&=\sum_{\alpha}p(\alpha|\mathbf{m})\rho_0^{(\alpha)}(\mathbf{x}')\\
&=\sum_{\mathbf{x}}\frac{P(\mathbf{m}|\mathbf{x})\rho_0(\mathbf{x})}{p(\mathbf{m})}\delta_{\mathbf{x},\mathbf{x}'}\\
&=\frac{P(\mathbf{m}|\mathbf{x})\rho_0(\mathbf{x}')}{p(\mathbf{m})}\\
\end{aligned}
\end{equation}
i.e., the posterior with respect to the initial microstate distribution $\rho_0$. Note that in this scenario, the initial average microstate distribution, given in general by Eq.~\ref{eq:mixture}, becomes identical to the microstate distribution itself:
\begin{equation}
\bar{\rho}_0=\sum_{\mathbf{x}_0}\rho_0(\mathbf{x}_0)\delta_{\mathbf{x}_0,\mathbf{x}}=\rho_0.
\end{equation}

\end{document}